\documentclass[%
 reprint,
 amsmath,amssymb,
 aps,
prstab,
]{revtex4-1}

\usepackage{graphicx}
\usepackage{dcolumn}
\usepackage{bm}
\usepackage{gensymb}
\usepackage{color}

\renewcommand{\Re}{{\rm Re}}
\renewcommand{\Im}{{\rm Im}}

\begin{document}


\title{Schottky Signal Modification as a Diagnostic Tool for Coherent Electron Cooling}

\author{W. F. Bergan}
\email[]{wbergan@bnl.gov}
\author{M. Blaskiewicz}

\affiliation{Brookhaven National Laboratory, Upton, New York 11973, USA}
\author{G. Stupakov}
\affiliation{SLAC National Accelerator Laboratory, Menlo Park, California 94025, USA}
\begin{abstract}
Coherent electron cooling is a promising technique to cool high-intensity hadron bunches by imprinting the noise in the hadron beam on a beam of electrons, amplifying the electron density modulations, and using them to apply cooling kicks to the hadrons. The typical size for these perturbations can be on the $\mu$m scale, allowing us to extend the reach of classical stochastic cooling by several orders of magnitude. However, it is crucial to ensure that the electron and hadron beams are longitudinally aligned within this same $\mu$m scale. In order to provide fast feedback for this process, we discuss the extension of signal suppression to coherent electron cooling, and show in both theory and simulation that certain components of the spectral noise in the hadron beam will be predictably modified at the several percent level, which may be detected by observations of the radiation of the hadron beam.

\end{abstract}

\maketitle

\section{Introduction}\label{sec:intro}
In high-intensity hadron storage rings, intrabeam scattering (IBS) and beam-beam effects will degrade the beam emittance over the length of the store, limiting machine luminosity. In particular, at the planned Electron-Ion Collider (EIC), the IBS times are expected to be at the timescale of a couple hours, and so some form of strong hadron cooling is necessary to achieve the physics goals \cite{cite:eic_cdr}.

The proposed method in this case is microbunched electron cooling (MBEC), a particular form of coherent electron cooling (CeC). This was first introduced in \cite{cite:ratner}, and the theory has since been developed extensively in \cite{cite:stupakov_initial, cite:stupakov_amplifier, cite:stupakov_transverse, cite:stupakov_fourth}. The premise of MBEC is that the hadron beam to be cooled copropagates with an electron beam in a straight ``modulator'' section, during which time the hadrons will provide energy kicks to the electrons. The two beams are then separated, and the electrons pass through a series of amplifiers to change this initial energy modulation into a density modulation and amplify it. The hadrons travel through their own chicane before meeting the electrons again in a straight ``kicker'' section. Here, the amplified density modulations in the electron beam provide energy kicks to the hadrons. By tuning the hadron chicane so that the hadron delay in travelling from the modulator to the kicker is energy-dependent, we may arrange it so that the energy kick that the hadron receives in the kicker tends to correct initial energy offsets. If the chicane also gives the hadrons a delay dependent on their transverse phase-space coordinates and if there is non-zero dispersion in the kicker, then the transverse emittance of the hadron beam can also be cooled. In the current design of an MBEC Cooler for the EIC, the typical scale of the electron density modulations at the top energy will be $\sim1\mu$m \cite{cite:eic_mbec_design}. This corresponds to about 4 orders of magnitude higher bandwidth than can be achieved with microwave stochastic cooling \cite{cite:microwave_stochastic_cooling, cite:stochastic_cooling_rhic}, allowing the cooling of dense hadron bunches, but also making alignment a challenge.

It is important that the hadron arrives in the kicker at the same time as the density perturbations which it had induced in the electron beam, or else it will receive entirely uncorrelated energy kicks \cite{cite:Seletskiy_2021zrn}. Comparing the $\sim100$m distance between modulator and kicker to the $\sim1\mu$m density perturbations in the electron beam, we see that the transit times of the electrons and hadrons must be maintained at a level of ten parts per billion. In order to commission and operate such a system, it is necessary to have some way to quickly measure the relative alignment of the electron and hadron beams at the sub-micron scale. Directly observing cooling would require waits on the timescale of hours, which would make commissioning painful and prevent any sort of fast feedback during operations. The method proposed here is to make use of ``signal modification,'' an extension of the well-known signal suppression of microwave stochastic cooling \cite{cite:sigsup_original, cite:sigsup_bisognano, cite:sigsup_ruggiero, cite:sigsup_sc} to the case of MBEC cooling. The principle of this method is that after the hadron beam has received its cooling kicks, it will propagate to a ``detector'' where the power of the hadron beam at particular wavelengths may be measured. If the hadron beam is well-aligned with the electron beam, this will produce a predictable change in the spectral content of the hadron beam on a single-pass basis.

\begin{figure}[!htbp]
\begin{center}
\includegraphics[width=1.0\columnwidth]{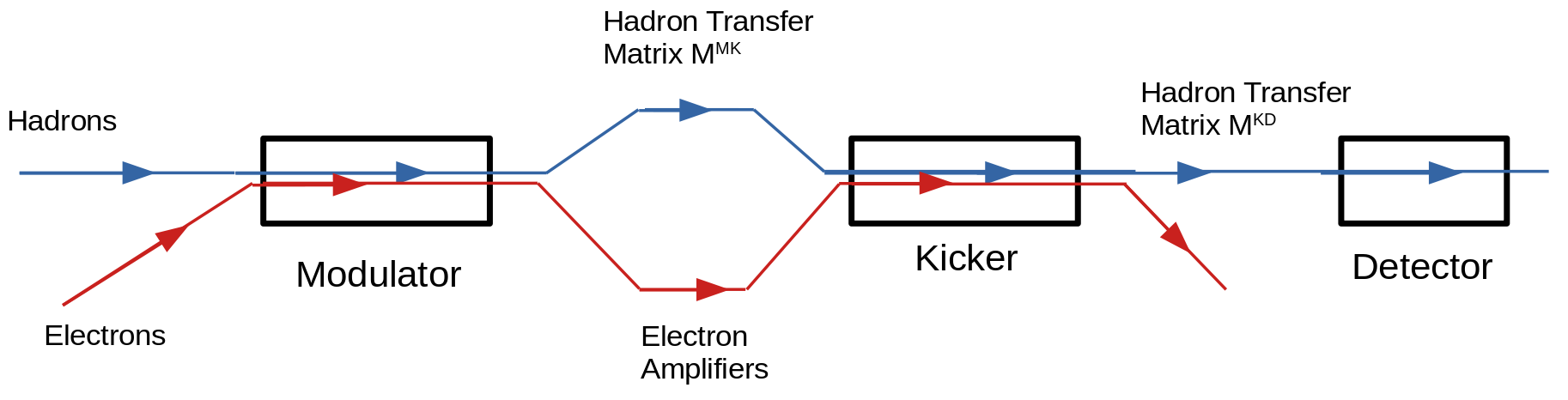}
\end{center}
\caption{\label{fig:layout} A schematic of the MBEC cooling section, including the usual modulator and kicker necessary for cooling, as well as the diagnostic detector where signal modification may be observed.}
\end{figure}

As our model, we assume that the cooling section consists of a modulator, where the hadron beam imprints on the electrons; a kicker, where the electrons provide an energy kick back to the hadrons; and a detector, where we will observe the density spectrum in the hadron beam. See Fig.~\ref{fig:layout}.

In Section \ref{sec:thry}, we provide a theoretical derivation of signal modification. In Section \ref{sec:sim}, we discuss simulation tools to model this process, and find good agreement with the theoretical predictions. In Section \ref{sec:detection}, we discuss what will be needed to measure such a signal experimentally. We present our outlook in Section \ref{sec:conclude} and conclude.

Note that, although this paper focuses on MBEC, the general form of these results will hold for coherent electron cooling with other amplification mechanisms.

\section{Theory}\label{sec:thry}

We derive here the theory of signal modification by directly propagating the particles themselves from the modulator to the detector with arbitrary 6-dimensional transfer matrices. Alternative derivations using the longitudinal Vlasov equation are presented in Appendix \ref{app:alternative}. In subsection \ref{subsec:thry_decohere}, we comment on decoherence of the signal when observing a range of frequencies and in subsection \ref{subsec:multi_kick}, we discuss an extension where the kicks are distributed throughout the length of the kicker.

We use phase-space coordinates $x$, $x'$, $y$, $y'$, $z$, and $\eta$, where the first four are the transverse positions and angles, $z$ is the particle's longitudinal position in the bunch, with positive $z$ values corresponding to the head of the bunch, and $\eta$ is the fractional energy offset of the particle.

In order to characterize the cooling process, we follow \cite{cite:stupakov_initial} and define a longitudinal wake function {\color{blue} in the 1D approximation}, such that the energy kick a hadron receives in the kicker is the convolution of this wake function with the longitudinal distribution of hadrons in the modulator. Explicitly, the fractional energy kick $\Delta \eta$ received by a hadron at longitudinal position $z$ within the bunch at the kicker is given by
\begin{align}\label{eqtn:wake_def}
\Delta\eta(z) = \frac{q^2}{E_0} \int_{-\infty}^{\infty}w(z+\Delta z-z')n(z') dz'
\end{align}
\noindent
where $q$ is the hadron charge, $E_0$ is the nominal hadron energy, $n(z)$ is the longitudinal hadron density in the modulator, and $\Delta z$ is the difference in modulator-to-kicker longitudinal delay between the hadron and electron beams. We also identify a corresponding impedance
\begin{align}\label{eqtn:imped_def}
    Z(k) = -\frac{1}{c}\int_{-\infty}^{\infty} w(z) e^{-ikz} dz
\end{align}

In order for these simple longitudinal wakes to hold, we require coherent oscillation of the beams, and therefore the electric field at our frequency of interest must be large compared to the beam sizes. In particular, we require \cite{cite:rev_ref1}:

\begin{align}\label{eqtn:1D}
    \Sigma_{\perp} \lesssim \frac{\beta\gamma\lambda}{2}
\end{align}
\noindent
where $\Sigma_{\perp}$ is the transverse beam size, $\beta$ is the relativistic beta, $\gamma$ is the relativistic gamma, and $\lambda$ is the beam oscillation wavelength. Examining the values in Tab.~\ref{tab:param}, we see that the largest beam size is that of the protons in the modulator and kicker (0.95mm), and the gamma factor is $293$, so our approximation is good for wavelengths above $\sim6.5\mu$m. As we will see, we will be interested mainly in wavelengths of $6 \mu$m, at the limit of applicability of the above. However, we note that the requirement of Eq.~\ref{eqtn:1D} is likely not a strict one---as it was shown in \cite{stupakov:2019-3D_cooling}, in the case when there is not amplification, the 3D theory gives exactly the same result as the 1D model which we employ here.

In order for us to neglect Landau damping in the electron beam amplifiers, one usually requires \cite{cite:rev_ref2}:

\begin{align}\label{eqtn:landau_damp}
    k \ll k_D,
\end{align}
\noindent
where $k$ is the wavenumber of our signal, $k_D \equiv \omega_p/c\sigma_{\beta}$ is the Debye wavenumber, and $\sigma_{\beta}$ is the spread in relativistic beta. This electron beam has $k_D = 4\times10^6$m$^{-1}$ in the amplifiers, larger than the $k = 10^6$m$^{-1}$ of our microbunching. One also has to keep in mind that the concept of Landau damping, with the condition of Eq.~\ref{eqtn:landau_damp}, is only valid asymptotically, after many plasma oscillations, which is not the case of the microbunching coherent cooling where the electron beam executes $\frac{1}{4}$ of the  plasma oscillation in the amplifier sections. A more accurate analysis of when one can ignore the Landau damping for this case was carried out in Section IV of \cite{cite:stupakov_amplifier}.

In order to ensure that the microbunching is not washed out by the energy spread and divergence of the beam within any of the straight sections, we require \cite{cite:rev_ref2, PhysRevLett.102.154801}:

\begin{align}
    \lambda \gg \frac{2L}{\beta^3\gamma^2}\sigma_{\eta}
\end{align}

and

\begin{align}
    \lambda \gg \frac{L}{\beta^2}\sigma_{x'}^2
\end{align}
\noindent
where $\sigma_{\eta}$ is the fractional energy spread of the beam, $\sigma_{x'}$ is the beam divergence, and $L$ is the length of a drift. Comparison with the parameters in Tab.~\ref{tab:param} shows that these are met for both the electrons and hadrons in all the straight sections.

We assume that there is no transverse correlation in the electron or hadron beam. Further discussion of the effects which this might have is available in \cite{cite:rev_ref4}.

We focus our attention on the peak region of the electron and hadron beams and take the limit of a longitudinally infinite and uniform plasma. Since the typical wake wavelength is on the order of a few microns, and the typical bunch lengths are a few mm or longer, this is a reasonable approximation. Finally, we assume that the hadron beam enters the modulator with no correlated structures on the scale of the wake wavelength. Although such structures will be generated within the kicker, their characteristic size is on the micron scale, much less than the millimeter scale longitudinal motion per turn, washing out any memory of the kick by the time the beam enters the modulator again.

As illustrated in Fig.~\ref{fig:layout}, we take the hadrons to have transfer matrix $M^{MK}$ between modulator and kicker and $M^{KD}$ between kicker and detector, with the transfer matrix between modulator and detector given by $M^{MD}$.

We treat our particles as existing in a region of length $L$, much larger than any length scale associated with the wake function, and assume periodic boundary conditions, so that we may arbitrarily shift the limits of integration in our integrals. In this model, we also consider the full 6-dimensional evolution of the hadron beam and ignore any collective effects during beam transport except for the electron-hadron interactions characterized by the wake function, as discussed above. We write the evolution of a hadron's position between modulator and detector as

\begin{align}\label{eqtn:delays}
z_d^{(i)} = &z_m^{(i)} + M^{MD}_{5u}\vec{x}_u^{(i)}\\\nonumber
+ &M^{KD}_{56}\sum_j \frac{q^2}{E_0} w(z_m^{(i)} + M^{MK}_{5u}\vec{x}_u^{(i)} + \Delta z - z_m^{(j)})
\end{align}
\noindent
where $z_m^{(i)}$ is the longitudinal position of particle $i$ in the modulator, $z_d^{(i)}$ is its position within the detector, and $\vec{x}^{(i)}$ are the phase-space coordinates of particles $i$ in the modulator. We use the convention that the repeated ``$u$'' subscript refers to sums over the 5 phase-space coordinates \textit{excluding} the longitudinal position. The summation over $j$ is over all particles within the hadron beam. The first two terms of this equation describe the modulator to kicker beam evolution by a simple transfer matrix, while the final term gives the contribution to delay due to the extra energy kick our particle receives in the kicker due to the wakes of all particles in the beam, including itself.

At the detector, the longitudinal density of the hadron beam is
\begin{align}\label{eqtn:rho_d}
n(z) = \sum_i \delta(z - z_d^{(i)})
\end{align}
\noindent
with corresponding density in Fourier space
\begin{align}\label{eqtn:rho_tilde_d}
\tilde{n}(k) &= \int_{-\infty}^{\infty}\sum_i e^{-ikz} \delta(z - z_d^{(i)}) dz\\\nonumber
&=\sum_i e^{-ikz_d^{(i)}}
\end{align}

The power in the hadron spectrum at a given wavenumber is then given by
\begin{align}\label{eqtn:pwr}
|\tilde{n}(k)|^2 = &\sum_{i,a}e^{-ik\big[z_d^{(i)} - z_d^{(a)}\big]}\\\nonumber
=N + &\sum_{i \neq a}e^{-ik\big[z_m^{(i)} - z_m^{(a)} + M^{MD}_{5u}\big(\vec{x}_u^{(i)} - \vec{x}_u^{(a)}\big)\big]}\\\nonumber
&\times e^{-ikM^{KD}_{56}q^2/E_0\sum_j w\big(z_m^{(i)} + M^{MK}_{5u}\vec{x}_u^{(i)} + \Delta z - z_m^{(j)}\big)}\\\nonumber
&\times e^{ikM^{KD}_{56}q^2/E_0\sum_j w\big(z_m^{(a)} + M^{MK}_{5u}\vec{x}_u^{(a)} + \Delta z - z_m^{(j)}\big)}
\end{align}
\noindent
where we have substituted in the expression for $z_d^{(i)}$ given in Eq.~\ref{eqtn:delays} and used the fact that the $i=a$ terms in the sum are all equal to $1$, giving us the $N$ out front, where $N$ is the number of particles in the length-$L$ section of the beam.

Typically, the kick from the wake is small, and so we may Taylor-expand the final two exponentials above to linear order. We thereby obtain

\begin{align}\label{eqtn:pwr_taylor}
|\tilde{n}(k)|^2 &= N + \sum_{i \neq a}e^{-ik\big[z_m^{(i)} - z_m^{(a)} + M^{MD}_{5u}\big(\vec{x}_u^{(i)} - \vec{x}_u^{(a)}\big)\big]}\\\nonumber
\times \big[1 &- ikM^{KD}_{56}\frac{q^2}{E_0}\sum_j w\big(z_m^{(i)} + M^{MK}_{5u}\vec{x}_u^{(i)} + \Delta z - z_m^{(j)}\big)\\\nonumber
&+ ikM^{KD}_{56}\frac{q^2}{E_0}\sum_j w\big(z_m^{(a)} + M^{MK}_{5u}\vec{x}_u^{(a)} + \Delta z - z_m^{(j)}\big)\big]
\end{align}
\noindent
with the effect of second-order terms considered in Appendix \ref{app:second_order}.

We now wish to take the expectation value of the beam power, requiring integrals over the 12 phase space coordinates of particles $i$ and $a$ and the longitudinal position of particle $j$. However, note that the dependence on $z_m^{(j)}$ appears only in the argument of the wake functions. If the total integral of the wake is zero, then, if particle $j$ is distinct from both particles $i$ and $a$, this integral will evaluate to 0. We then need only consider the terms $j=i$ and $j=a$ in those sums. The beam power can then be written as

\begin{align}\label{eqtn:pwr_taylor2}
|\tilde{n}(k)|^2 &= N + \sum_{i \neq a}e^{-ik\big[z_{ia} + M^{MD}_{5u}\big(\vec{x}_u^{(i)} - \vec{x}_u^{(a)}\big)\big]}\\\nonumber
\times [1 &- ikM^{KD}_{56}\frac{q^2}{E_0} w(M^{MK}_{5u}\vec{x}_u^{(i)} + \Delta z)\\\nonumber
&+ ikM^{KD}_{56} \frac{q^2}{E_0}w(M^{MK}_{5u}\vec{x}_u^{(a)} + \Delta z)\\\nonumber
&- ikM^{KD}_{56} \frac{q^2}{E_0}w(z_{ia} + M^{MK}_{5u}\vec{x}_u^{(i)} + \Delta z)\\\nonumber
&+ ikM^{KD}_{56} \frac{q^2}{E_0}w(-z_{ia} + M^{MK}_{5u}\vec{x}_u^{(a)} + \Delta z)]
\end{align}
\noindent
where we have made the definition $z_{ia} \equiv z_m^{(i)} - z_m^{(a)}$. Since we assume a homogeneous hadron bunch, $z_m^{(i)}$ and $z_m^{(a)}$ themselves are irrelevant and $z_{ia}$ has the same probability distribution as them.

We note in the above formula that the first three terms have their only $z_{ia}$ dependence in the leading exponential, and so performing an average over all $z_{ia}$ will be zero. We therefore need only focus on the 4th and 5th terms. Approximating the $N(N-1)$ terms in the above sum as $N^2$, the relevant integral for the 4th term is 

\begin{align}\label{eqtn:4th_term}
&-N^2ikM_{56}^{KD} \int_{-L/2}^{L/2}  dz_{ia}/L \int_{-\infty}^{\infty} d^5\vec{x}^{(i)} d^5\vec{x}^{(a)} \rho\big(\vec{x}^{(i)}\big) \rho\big(\vec{x}^{(a)}\big)\\\nonumber
&\times \frac{q^2}{E_0}w(z_{ia} + M^{MK}_{5u}\vec{x}_u^{(i)} + \Delta z)
e^{-ik\big[z_{ia} + M^{MD}_{5u}\big(\vec{x}_u^{(i)} - \vec{x}_u^{(a)}\big)\big]}
\end{align}
\noindent
where $\rho\big(\vec{x}\big)$ the hadron phase-space density in the modulator within the 5 phase-space coordinates excluding longitudinal position.

Approximating the longitudinal integral as extending from $-\infty$ to $\infty$, using the impedance of Eq.~\ref{eqtn:imped_def}, and making an appropriate change of variables to $z' \equiv  z_{ia} + M^{MK}_{5u}\vec{x}_u^{(i)} + \Delta z$, the longitudinal integral in Eq.~\ref{eqtn:4th_term} may be evaluated, leaving us with

\begin{align}\label{eqtn:4th_term_2}
&N^2ikM_{56}^{KD} \frac{q^2c}{E_0}\frac{1}{L} Z(k) e^{ik\Delta z}\\\nonumber
&\times\int_{-\infty}^{\infty} d^5\vec{x}^{(i)} d^5\vec{x}^{(a)} \rho\big(\vec{x}^{(i)}\big) \rho\big(\vec{x}^{(a)}\big)\\\nonumber
&\times e^{-ik\big[M^{MD}_{5u}\big(\vec{x}_u^{(i)} - \vec{x}_u^{(a)}\big) - M^{MK}_{5u}\vec{x}_u^{(i)}\big]}
\end{align}

To perform the remaining integrals, we write the evolution of the phase-space coordinates explicitly in terms of action-angle variables and Courant-Snyder parameters at the start of the transfer matrix \cite{cite:sy_lee}, finding

\begin{align}\label{eqtn:action_angle}
M_{5u}\vec{x}_u &= M_{51}(\sqrt{2J_x\beta_x}\cos(\phi_x)+D_x\eta)\\\nonumber
&+ M_{52}(-\sqrt{2J_x/\beta_x}[\sin(\phi_x)+\alpha_x\cos(\phi_x)]+D'_x\eta)\\\nonumber
&+ M_{53}(\sqrt{2J_y\beta_y}\cos(\phi_y)+D_y\eta)\\\nonumber
&+ M_{54}(-\sqrt{2J_y/\beta_y}[\sin(\phi_y)+\alpha_y\cos(\phi_y)]+D'_y\eta)\\\nonumber
&+ M_{56}\eta\\\nonumber
&\\\nonumber
&= (M_{51}-\frac{\alpha_x}{\beta_x}M_{52})\sqrt{2J_x\beta_x}\cos(\phi_x)\\\nonumber
&- M_{52}\sqrt{2J_x/\beta_x}\sin(\phi_x)\\\nonumber
&+ (M_{53}-\frac{\alpha_y}{\beta_y}M_{54})\sqrt{2J_y\beta_y}\cos(\phi_y)\\\nonumber
&- M_{54}\sqrt{2J_y/\beta_y}\sin(\phi_y)\\\nonumber
&+ (D_xM_{51} + D'_xM_{52} + D_yM_{53} + D'_yM_{54} + M_{56})\eta\\\nonumber
&\\\nonumber
\equiv & \hat{M}_{51}\hat{x} + \hat{M}_{52}\hat{x}' + \hat{M}_{53}\hat{y} + \hat{M}_{54}\hat{y}' + \hat{M}_{56}\hat{\eta}
\end{align}
\noindent
with
\begin{align}\label{eqtn:m_tilde}
&\hat{M}_{51} \equiv M_{51}-\frac{\alpha_x}{\beta_x}M_{52}\\\nonumber
&\hat{M}_{52} \equiv M_{52}\\\nonumber
&\hat{M}_{53} \equiv M_{53}-\frac{\alpha_y}{\beta_y}M_{54}\\\nonumber
&\hat{M}_{54} \equiv M_{54}\\\nonumber
&\hat{M}_{56} \equiv D_xM_{51} + D'_xM_{52} + D_yM_{53} + D'_yM_{54} + M_{56}
\end{align}

For a Gaussian beam, the $\hat{x}$, $\hat{x}'$, $\hat{y}$, $\hat{y}'$, and $\hat{\eta}$ are normally distributed with 
\begin{align}\label{eqtn:sigmas}
        &\sigma_{\hat{x}} = \sqrt{\epsilon_x\beta_x}\\\nonumber
        &\sigma_{\hat{x}'} = \sqrt{\epsilon_x/\beta_x}\\\nonumber
        &\sigma_{\hat{y}} = \sqrt{\epsilon_y\beta_y}\\\nonumber
        &\sigma_{\hat{y}'} = \sqrt{\epsilon_y/\beta_y}\\\nonumber
        &\sigma_{\hat{\eta}} = \sigma_{\eta}
\end{align}
\noindent
where the $\epsilon$ are the horizontal and vertical emittances and $\sigma_{\eta}$ is the fractional energy spread.
In this case, the remaining ten integrals in Eq.~\ref{eqtn:4th_term_2} may be performed, yielding

\begin{align}\label{eqtn:4th_term_final}
&ikM_{56}^{KD}\frac{q^2c}{E_0} \frac{N^2}{L} Z(k) e^{ik\Delta z}\\\nonumber
&\times e^{-\frac{k^2}{2} \sum_{u\neq5} \sigma^2_{\hat{x}_u}\Big[\big(\hat{M}^{MD}_{5u}\big)^2 + \big(\hat{M}^{MD}_{5u} - \hat{M}^{MK}_{5u}\big)^2\Big]}
\end{align}

A similar procedure may be applied to the fifth term of Eq.~\ref{eqtn:pwr_taylor2}, resulting in 

\begin{align}\label{eqtn:5th_term_final}
&-ikM_{56}^{KD} \frac{q^2c}{E_0}\frac{N^2}{L} Z(-k) e^{-ik\Delta z}\\\nonumber
&\times e^{-\frac{k^2}{2} \sum_{u\neq5} \sigma^2_{\hat{x}_u}\Big[\big(\hat{M}^{MD}_{5u}\big)^2 + \big(\hat{M}^{MD}_{5u} - \hat{M}^{MK}_{5u}\big)^2\Big]}
\end{align}

Making use of the fact that, for real wakes, $Z(-k) = Z^*(k)$, and defining $n_0 \equiv N/L$ as the mean linear density of the hadrons, we may sum Eqs.~\ref{eqtn:4th_term_final} and \ref{eqtn:5th_term_final} and incorporate them back into Eq.~\ref{eqtn:pwr_taylor2}, obtaining

\begin{align}\label{eqtn:pwr_final}
|\tilde{n}(k)|^2 &= N - 2Nn_0\frac{q^2c}{E_0}kM_{56}^{KD}\\\nonumber
&\times[\Re(Z(k))\sin(k\Delta z) + \Im(Z(k))\cos(k\Delta z)]\\\nonumber
&\times e^{-\frac{k^2}{2} \sum_{u\neq5} \sigma^2_{\hat{x}_u}\Big[\big(\hat{M}^{MD}_{5u}\big)^2 + \big(\hat{M}^{MD}_{5u} - \hat{M}^{MK}_{5u}\big)^2\Big]}
\end{align}

Then, the fractional change in beam power as a function of electron/hadron misalignment is
\begin{align}\label{eqtn:pwr_relative}
\frac{\Delta |\tilde{n}(k)|^2}{|\tilde{n}(k)|^2} &= -2n_0\frac{q^2c}{E_0}kM_{56}^{KD}\\\nonumber
&\times[\Re(Z(k))\sin(k\Delta z) + \Im(Z(k))\cos(k\Delta z)]\\\nonumber
&\times e^{-\frac{k^2}{2} \sum_{u\neq5} \sigma^2_{\hat{x}_u}\Big[\big(\hat{M}^{MD}_{5u}\big)^2 + \big(\hat{M}^{MD}_{5u} - \hat{M}^{MK}_{5u}\big)^2\Big]}
\end{align}

Note that if $\Re(Z(k)) = 0$ and $\Im(Z(k)) > 0$ (which is the case considered below, see Eq.~\ref{eqtn:imped_approx}), and if $M_{56}^{KD}>0$, then perfect alignment of the hadrons and electrons, ie $\Delta z = 0$, results in $\Delta |\tilde{n}(k)|^2 < 0,$ corresponding to noise suppression below the shot noise in the beam. Such noise suppression has been previously studied theoretically in \cite{PhysRevLett.102.154801,PhysRevSTAB.14.060710} and observed experimentally in \cite{PhysRevLett.109.034801,Gover:2012ts,cite:rev_ref5}. Our result Eq.~\ref{eqtn:pwr_relative} is an agreement with the theoretical analysis of \cite{PhysRevSTAB.14.060710}.

\subsection{Decoherence}\label{subsec:thry_decohere}

The above posits that the amount of signal modification has a purely sinusoidal dependence on the electron/hadron misalignment, ie,
\begin{align}\label{eqtn:decoherence_ideal}
    \frac{\Delta |\tilde{n}(k)|^2}{|\tilde{n}(k)|^2} = A(k) \cos(k\Delta z + \theta_0)
\end{align}
\noindent
where $A$ is some amplitude and $\theta_0$ is a phase, equal to $0$ for an antisymmetric wake.

However, the above derivation assumes an observation at a pure single frequency. If we take the more realistic case that we sample some range of frequencies with bandwidth $\Delta k$, the signal amplitude will be
\begin{align}\label{eqtn:decoherence_true}
    \frac{\Delta |\tilde{n}(k)|^2}{|\tilde{n}(k)|^2} &\approx A(k)\frac{1}{\Delta k}\int_{k-\Delta k/2}^{k+\Delta k/2} \cos(k'\Delta z + \theta_0) dk'\\\nonumber
    &= A(k)\cos(k\Delta z + \theta_0) \frac{\sin(\Delta k\Delta z/2)}{\Delta k\Delta z/2}
\end{align}
\noindent
so that the amplitude of the oscillations will decay over lengths of $\sim 1/\Delta k$.

\subsection{Integrated Kicker}\label{subsec:multi_kick}

The above analysis assumes that the full kick to the hadrons takes place at a single point in the kicker. However, in the more realistic case where the kick is applied over the full length of the kicker, we may change Eq.~\ref{eqtn:delays} to
\begin{align}
    &z_d^{(i)} = z_m^{(i)} + M^{MD}_{5u}\vec{x}_u^{(i)}\\\nonumber
&+ \frac{1}{N_s}\sum_{s=1}^{N_s}M^{K_sD}_{56}\sum_j \frac{q^2}{E_0} w(z_m^{(i)} + M^{MK_s}_{5u}\vec{x}_u^{(i)} + \Delta z - z_m^{(j)})
\end{align}
\noindent
where we now split our kick in the kicker into $N_s$ fractional kicks at locations $K_s$.

The rest of the analysis may be carried through in the same way as before, and we arrive at

\begin{align}
    \frac{\Delta |\tilde{n}(k)|^2}{|\tilde{n}(k)|^2} &= -2n_0\frac{q^2c}{E_0}\frac{1}{N_s}\sum_{s=1}^{N_s}kM_{56}^{K_sD}\\\nonumber
&\times[\Re(Z(k))\sin(k\Delta z) + \Im(Z(k))\cos(k\Delta z)]\\\nonumber
&\times e^{-\frac{k^2}{2} \sum_{u\neq5} \sigma^2_{\hat{x}_u}\Big[\big(\hat{M}^{MD}_{5u}\big)^2 + \big(\hat{M}^{MD}_{5u} - \hat{M}^{MK_s}_{5u}\big)^2\Big]}
\end{align}

However, we have not observed any significant difference numerically between using the above equation and Eq.~\ref{eqtn:pwr_relative}, with the single kick taken at the kicker center.

\section{Simulation}\label{sec:sim}

In order to check the validity and limits of the above theory, we make use of simulation. We first examine the case of a perfectly linear simulation, where the fractional energy kick to a given hadron in the kicker is simply the convolution of the wake function with the longitudinal hadron distribution in the modulator. We then turn our attention to a more detailed simulation, where the electrons and hadrons are tracked with a particle-in-cell code in order to incorporate saturation effects. In this and future sections, we consider the cooling parameters currently planned for 275 GeV protons in the EIC, listed in Tab.~\ref{tab:param}. The electron optics are assumed to be kept roughly constant within the modulator, kicker, and amplifiers through the use of focusing. Due to their much higher energy, the hadrons see the modulator and kicker as drifts, with the optics parameters specified at the center. Electron chicanes are defined from the end of one straight section to the start of the next one, while the proton $M_{56}$ and phase advances are evaluated from modulator center to kicker center.

\begin{table*}[!hbt]
   \centering
   \caption{Parameters for Longitudinal and Transverse Cooling}
   \begin{tabular}{lc}
           \textit{\textbf{Geometry}}                                           &                      \\
           Modulator Length (m)                                                 & 45                   \\ 
           Kicker Length (m)                                                    & 45                   \\ 
           Number of Amplifier Straights                                        & 2                    \\ 
           Amplifier Straight Lengths (m)                                       & 37                   \\ 
           \textit{\textbf{Proton Parameters}}                                  &                      \\
           Energy (GeV)                                                         & 275                  \\
           Protons per Bunch                                                    & 6.9e10               \\
           Average Current (A)                                                  & 1                    \\
           Proton Bunch Length (cm)                                             & 6                    \\ 
           Proton Fractional Energy Spread                                      & 6.8e-4               \\ 
           Proton Emittance (x/y) (nm)                                          & 11.3 / 1             \\ 
           Horizontal/Vertical Proton Betas in Modulator (m)                    & 39 / 39              \\ 
           Horizontal/Vertical Proton Dispersion in Modulator (m)               & 1 / 0                \\ 
           Horizontal/Vertical Proton Dispersion Derivative in Modulator        & -0.023 / 0           \\ 
           Horizontal/Vertical Proton Betas in Kicker (m)                       & 39 / 39              \\ 
           Horizontal/Vertical Proton Dispersion in Kicker (m)                  & 1 / 0                \\ 
           Horizontal/Vertical Proton Dispersion Derivative in Kicker           & 0.023 / 0            \\ 
           Proton Horizontal/Vertical Phase Advance (rad)                       & 4.79 / 4.79          \\ 
           Proton M56 between Centers of Modulator and Kicker (mm)              & -2.26                \\ 
           \textit{\textbf{Electron Parameters}}                                &                      \\
           Energy (MeV)                                                         & 150                  \\
           Electron Bunch Charge (nC)                                           & 1                    \\ 
           Electron Bunch Length (mm)                                           & 7                    \\ 
           Electron Peak Current (A)                                            & 17                   \\ 
           Electron Fractional Slice Energy Spread                              & 1e-4                 \\ 
           Electron Normalized Emittance (x/y) (mm-mrad)                        & 2.8 / 2.8            \\ 
           Horizontal/Vertical Electron Betas in Modulator (m)                  & 40 / 20              \\ 
           Horizontal/Vertical Electron Betas in Kicker (m)                     & 4 / 4                \\ 
           Horizontal/ Vertical Electron Betas in Amplifiers (m)                & 1 / 1                \\ 
           M56 in First Electron Chicane (mm)                                   & 5                    \\ 
           M56 in Second Electron Chicane (mm)                                  & 5                    \\ 
           M56 in Third Electron Chicane (mm)                                   & -11                  \\ 
           \textit{\textbf{Cooling Times}}                                      &                      \\
           Horizontal/Vertical/Longitudinal IBS Times (hours)                   & 2.0 / - / 2.9        \\ 
           Horizontal/Vertical/Longitudinal Cooling Times (hours)               & 1.8 / - / 2.8        \\ 
   \end{tabular}
   \label{tab:param}
\end{table*}

\subsection{Linear Simulation}\label{subsec:sim_lin}

We simulate a 50$\mu$m length of the hadron beam at peak electron and hadron beam currents. Since the bunch lengths of the hadron and electron beams are a few cm and mm, respectively, we may assume constant longitudinal beam densities in our region of interest, and take periodic boundary conditions. We start by seeding 1 million hadron macroparticles in the modulator, representing 23 million real hadrons. In order to match the noise statistics in the real beam at both the modulator and detector, we perform the seeding with sub-Poisson noise, as follows. We create a 2-dimensional grid in the hadron phase space. One coordinate represents the hadron's longitudinal position at the center of the modulator, and the other represents its delay in travelling from the modulator center to the detector. Using the notation of Eq.~\ref{eqtn:m_tilde}, this delay may be expressed as \cite{cite:icfa_lebedev}

\begin{align}\label{eqtn:rms_delay}
\sigma^2_{\Delta z} = &\hat{M}_{51}^2\beta_x\epsilon_x + \hat{M}_{52}^2\epsilon_x/\beta_x\\\nonumber
+&\hat{M}_{53}^2\beta_y\epsilon_y + \hat{M}_{54}^2\epsilon_y/\beta_y\\\nonumber
+ &\hat{M}_{56}^2\sigma_{\eta}^2
\end{align}

This grid extends in 32768 uniform steps over the 50$\mu$m length of beam to be simulated, and in 100 uniform steps between $\pm5\sigma_{\Delta z}$ in the hadron delay. We assume that the hadrons are distributed uniformly longitudinally and that they have a Gaussian distribution (mean 0, standard deviation $\sigma_{\Delta z}$) with respect to the delay. Multiplying this distribution by the total number of hadrons expected within a 50$\mu$m slice of beam, we arrive at the expected number of hadrons in each bin. We then add on appropriate pseudo-random Poisson noise to obtain the number of real hadrons in each bin.

To assign the macroparticles, we iterate over all the bins, with the loop over position nested inside the loop over delays. We assign each bin in turn a number of macroparticles equal to the number of real hadrons in that bin divided by the number of real hadrons represented by each macroparticle. Each macroparticle is given a position and delay distributed uniformly within the bin. Underflow and overflow are carried to the next bin. Each macroparticle is assigned a horizontal angle (a pseudo-random number between $0$ and $2\pi$) and a horizontal action (a pseudo-random number drawn from an exponential distribution of mean $\epsilon_x$). Similar procedures are used to obtain the vertical action and angle. The fractional energy error is chosen to provide the hadron with its assigned delay. The actions, angles, and energy errors, together with the optics at the center of the modulator, allow the construction of the particle phase-space coordinates \cite{cite:sy_lee}.

In the case where the modulator-to-detector transfer matrix is an identity matrix, we assign the hadron fractional energy offset directly, replacing $\sigma_{\Delta z}$ in the above with $\sigma_{\eta}$.

\begin{figure}[!htbp]
\begin{center}
\includegraphics[width=1.0\columnwidth]{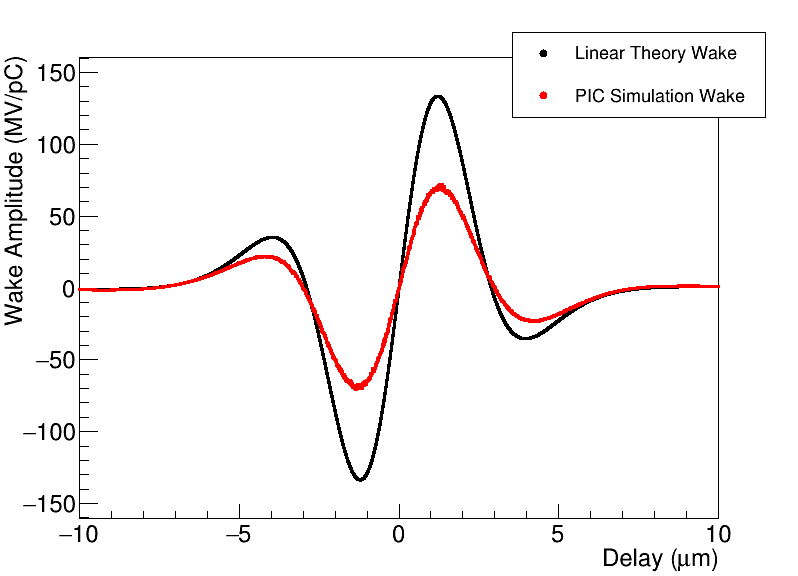}
\end{center}
\caption{\label{fig:wakes_both} Wake functions both from linear theory (without saturation) and simulated by our PIC code (with saturation).}
\end{figure}

We obtain the position of each hadron macroparticle in the kicker by using the modulator-to-kicker transfer matrix in the design optics, and save this pre-kicker distribution. We then apply a longitudinal shift to all hadrons equally, corresponding to a longitudinal electron-hadron misalignment. We assign an energy kick to each macroparticle by convolving the ideal wake function with the longitudinal hadron density distribution in the modulator. The ideal wake function is computed using the procedures of \cite{cite:stupakov_initial, cite:stupakov_amplifier, cite:stupakov_transverse} for the case of two amplification straights and unmatched Gaussian electron and hadron beams with arbitrary horizontal and vertical beam sizes. A plot of this idealized wake is shown in black in Fig.~\ref{fig:wakes_both}. Note that this wake is antisymmetiric, which means that $\Re(Z(k)) = 0$ and the first term in the square brackets on the right hand side of Eq.~\ref{eqtn:pwr_relative} vanishes. We then translate the hadrons to the detector element using a kicker-to-detector transfer matrix equal to the inverse of the modulator-to-kicker transfer matrix. We chose this matrix because we had found empirically that it gives a sizable signal and thereby reduces the impact of numerical noise. This will be further justified in subsection \ref{subsec:optimal_obs}. Finally, we perform a fast Fourier transform (FFT) on the hadron linear density distribution to obtain the amplitude of the spectrum. Squaring these values gives us the spectral power, and we compute the fractional change from the initial spectral power of the hadrons. We repeat this process from the saved pre-kicker distribution, but apply different longitudinal misalignments. Finally, we repeat this whole procedure from the beginning for 100 different random noise seeds and take the average fractional change in spectral power for each delay value.

\begin{figure}[!htbp]
\begin{center}
\includegraphics[width=1.0\columnwidth]{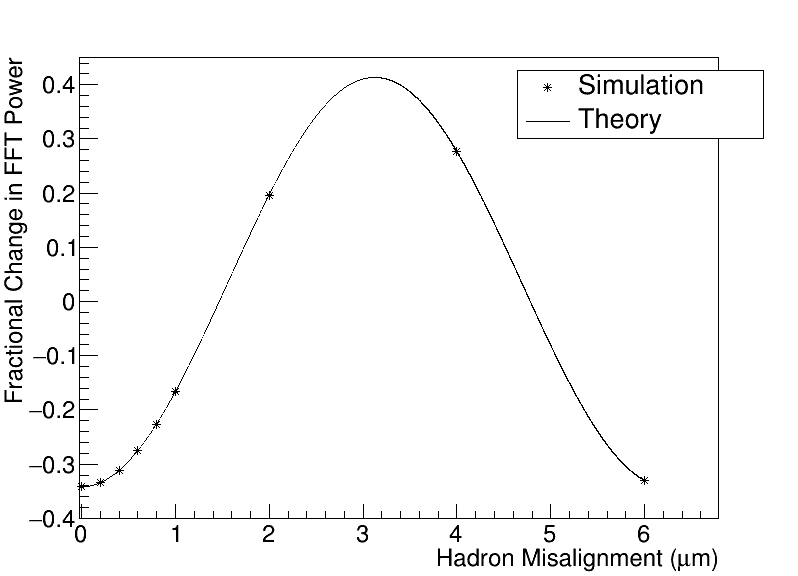}
\end{center}
\caption{\label{fig:sim_linear} Fractional change in the spectral power in the hadron beam at the $6.25\mu$m wavelength for the linear case. Excellent agreement between theory and simulation is observed.}
\end{figure}

We plot these simulated results and the theory prediction in Fig.~\ref{fig:sim_linear}, finding excellent agreement. Note that the theoretical wake has a constant vertical shift of $\sim 0.04$ due to higher-order terms, as discussed in Appendix \ref{app:second_order}.

\subsection{Nonlinear Simulation}\label{subsec:sim_nonlin}

As a more realistic model, we explicitly track the hadrons and electrons step-by-step through the modulator, kicker, and amplifiers. Much of the process for the hadrons is the same as in the linear case, described in subsection \ref{subsec:sim_lin}. The main difference is that, rather than simply convolve the wake with the hadron density distribution, we explicitly include the electron beam and model its interactions with itself and the hadrons in a custom particle-in-cell (PIC) code. This is described in \cite{cite:ipac2021_pic}, and a summary of its operation and changes is given below. The electrons are initialized analogously to the hadrons with 10 real electrons per macroparticle, but with only their longitudinal coordinates described. As such, the binning is done as a function of the electron longitudinal position and its fractional energy offset. The hadrons are initialized at the center of the modulator as described in subsection \ref{subsec:sim_lin} using the modulator optics of Tab.~\ref{tab:param}, then back-propagated to the start of the modulator so as to have the correct initial distribution. We track the electrons and hadrons through the modulator in 2 and 6 phase-space dimensions, respectively, using a simple kick-drift model. We model the inter-particle interactions using the disc model of \cite{cite:stupakov_initial, cite:stupakov_amplifier, cite:stupakov_transverse}, and convolve the associated force function with the hadron and electron longitudinal density distributions to obtain the kicks to each species. We then apply the transfer matrix of a short drift length to update the particle positions. This process is repeated through the length of the modulator. The electron and hadron chicanes are modelled as simple transfer matrices. In bringing the hadrons to the kicker, the modulator-to-kicker transfer matrix is multiplied by the inverse transfer matrices of half the modulator and kicker drifts, so as to have the correct transfer matrix between the element centers, while the electrons are explicitly tracked through the amplification section (3 chicanes and 2 straights), using the same PIC code. In the kicker, both species are again tracked with the PIC code to get accurate kicks to the hadrons. In bringing the hadrons to the detector, we multiply the kicker-to-detector transfer matrix by half the inverse transfer matrix of the kicker drift in order to have the correct transfer matrix from the kicker center to the detector. Using the transfer matrix from the center of the kicker is justified by the fact that this is the average location where the kick will take place.

\begin{figure}[!htbp]
\begin{center}
\includegraphics[width=1.0\columnwidth]{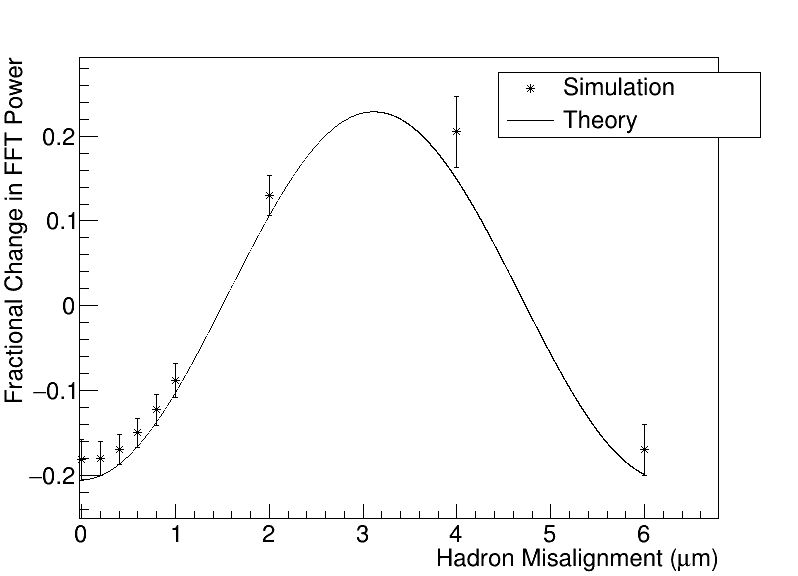}
\end{center}
\caption{\label{fig:sim_nonlinear} Fractional change in the spectral power in the hadron beam at the $6.25\mu$m wavelength for the fully nonlinear case. Good agreement between theory and simulation is observed.}
\end{figure}

In comparing to theory, we reduce the wake amplitude to account for saturation in the electron beam, as described in \cite{cite:ipac2021_pic}, and shown in red in Fig.~\ref{fig:wakes_both}. We incorporate the $\sim 0.01$ offset of Appendix \ref{app:second_order} into Eq.~\ref{eqtn:pwr_relative} and compare with the simulation results, as shown in Fig.~\ref{fig:sim_nonlinear}. Good agreement is observed.

\section{Detection of the Signal}\label{sec:detection}

So far, we have focused on the derivation and validation of the signal modification theory. However, for this to be useful, we must be able to physically detect it. We therefore look for the optimal kicker-to-detector transfer matrix and wavelength at which to observe the signal, and then examine the possibility of detecting the hadron beam density modulation in the radiation of an EIC dipole.

\subsection{Optimal Parameters}\label{subsec:optimal_obs}

\begin{figure}[!htbp]
\begin{center}
\includegraphics[width=1.0\columnwidth]{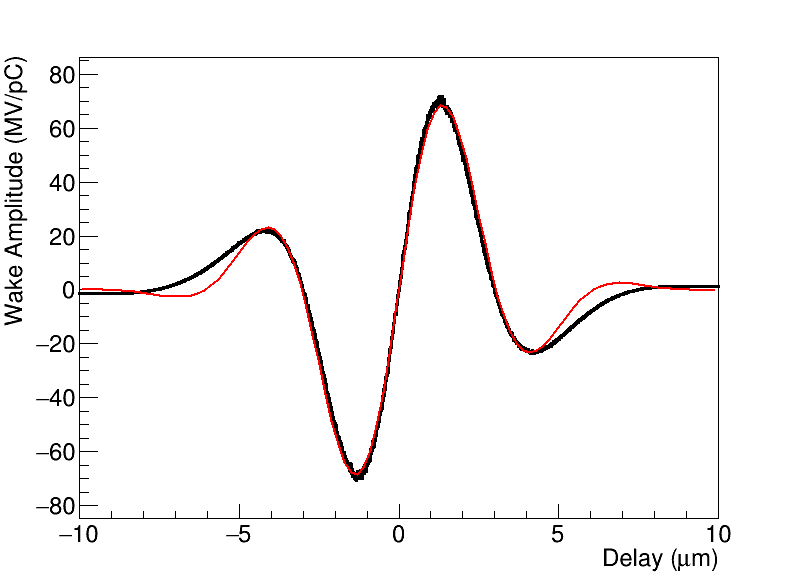}
\end{center}
\caption{\label{fig:wake_fit_sat} Fit of the simulated wake (with saturation) to a function of the form given in Eq.~\ref{eqtn:wake_approx}: $w(z) = A \sin(\kappa z)e^{-z^2/2\lambda^2}$. We find fit parameters of $A = 78.4$MV/pC, $\kappa = 1.031/\mu$m, and $\lambda = 2.768\mu$m, and see that this matches the data fairly well.}
\end{figure}

In order to perform the optimization, it is helpful to make use of a simplified, analytic form for the wake function. In \cite{cite:wake_approx}, it was proposed to approximate the wake using a simple fit function equal to a sine wave with Gaussian decay. Making slight changes to the parameters, we may write

\begin{align}\label{eqtn:wake_approx}
w(z) = A \sin(\kappa z)e^{-z^2/2\lambda^2}
\end{align}
\noindent
where $A$, $\kappa$, and $\lambda$ are fit parameters roughly corresponding to the wake amplitude, wavenumber, and falloff-distance, respectively. We find that this function fits our wake with saturation fairly well, as shown in Fig.~\ref{fig:wake_fit_sat}, and we arrive at values of $A = 78.4$MV/pC, $\kappa = 1.031/\mu$m, and $\lambda = 2.768\mu$m.

The corresponding impedance of the simplified wake of Eq.~\ref{eqtn:wake_approx} is given by 

\begin{align}\label{eqtn:imped_approx}
    Z(k) = -\frac{A\lambda i}{2c}\sqrt{2\pi}\Big[e^{-\lambda^2(k+\kappa)^2/2} - e^{-\lambda^2(k-\kappa)^2/2}\Big]
\end{align}

Putting the above into Eq.~\ref{eqtn:pwr_relative}, we obtain an expression for signal modification amplitude:

\begin{align}\label{eqtn:pwr_relative2}
\frac{\Delta |\tilde{n}^2(k)|^2}{|\tilde{n}^2(k)|^2} &= \frac{q^2}{E_0}A\lambda n_0kM_{56}^{KD}\sqrt{2\pi}\cos(k\Delta z)\\\nonumber
&\times \Big[e^{-\lambda^2(k+\kappa)^2/2} - e^{-\lambda^2(k-\kappa)^2/2}\Big]\\\nonumber
&\times e^{-\frac{k^2}{2} \sum_{u\neq5} \sigma^2_{\tilde{x}_u}\bigg[\Big(\tilde{M}^{MD}_{5u}\Big)^2 + \Big(\tilde{M}^{MK}_{5u} - \tilde{M}^{MD}_{5u}\Big)^2\bigg]}
\end{align}

Examining the above, we see that once we have fixed the cooling parameters, including the transfer matrix from the modulator to the kicker, the only variables which we can alter which will have any impact on the signal modification are the wavenumber of the density modulation we wish to observe and the $M_{5u}$ transfer elements from the kicker to the detector. (Note that these parameters alone are also sufficient to specify the $M_{5u}$ transfer matrix elements from the modulator to the detector.) If we have no vertical dispersion, $M^{MK}_{53} = M^{MK}_{54} = 0$, and it is easy to see from Eq.~\ref{eqtn:pwr_relative2} that the corresponding elements in the kicker-to-detector transfer matrix should also be equal to $0$. The magnitude of the signal amplitude in Eq.~\ref{eqtn:pwr_relative2} may be easily maximized with respect to $k$, $M^{KD}_{51}$, $M^{KD}_{52}$, and $M^{KD}_{56}$, obtaining values of $M_{51}^{KD} = -8.44\times 10^{-4}$, $M_{52}^{KD} = -3.00\times 10^{-2}$m, $M_{56}^{KD} = 2.36\times10^{-3}$m, and $k = 1.04 \times10^{6}/$m, with Eq.~\ref{eqtn:pwr_relative2} taking on the numerical value $\frac{\Delta |\tilde{n}^2(k)|^2}{|\tilde{n}^2(k)|^2} = -0.22\cos(k\Delta z)$. (For comparison, the simulations in Section \ref{sec:sim}, setting $M^{KD}$ to the inverse of $M^{MK}$ were made with near-optimal parameters $M_{51}^{KD} = -7.93\times 10^{-4}$, $M_{52}^{KD} = -2.86\times 10^{-2}$m, $M_{56}^{KD} = 2.26\times10^{-3}$m, and $k = 1.06 \times10^{6}/$m.)

\subsection{Drift Plus Dipole}\label{subsec:dipole_obs}

\begin{figure}[!htbp]
\begin{center}
\includegraphics[width=1.0\columnwidth]{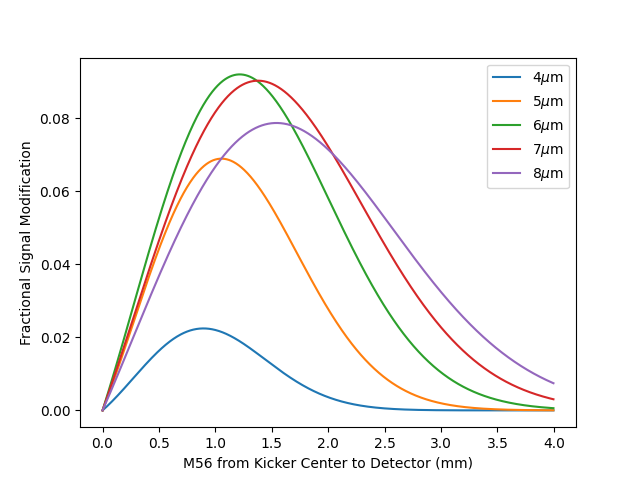}
\end{center}
\caption{\label{fig:scan_wavelength_discrete} Fractional signal modification as a function of $M_{56}$ between the kicker center and detection point for a variety of observation wavelengths. We find an optimal observation wavelength of $6\mu$m with an $M_{56}$ of 1.2mm. This would correspond to a total drift of over 100m, which would not be feasible in the available space.}
\end{figure}

\begin{figure}[!htbp]
\begin{center}
\includegraphics[width=1.0\columnwidth]{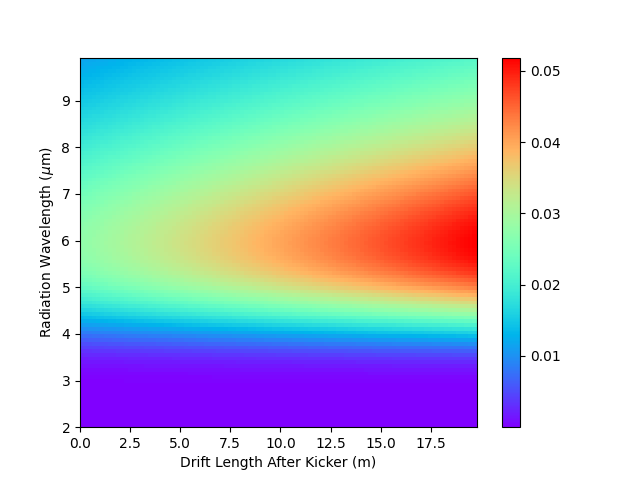}
\end{center}
\caption{\label{fig:scan_wavelength} Fractional signal modification expected at the start of the dipole if we observe the hadron density modulations of a specified wavelength after traversing a specified drift length after the kicker. We find an optimal observation wavelength of roughly  $6\mu$m over a range of drift lengths, giving a few percent signal modification amplitude.}
\end{figure}

\begin{figure}[!htbp]
\begin{center}
\includegraphics[width=1.0\columnwidth]{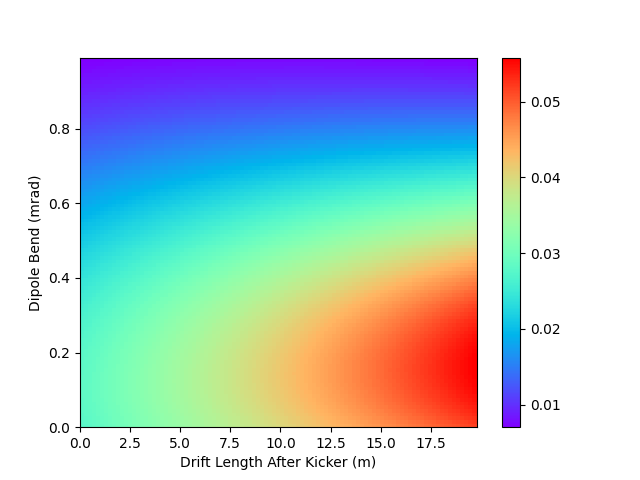}
\end{center}
\caption{\label{fig:scan_dipole} Fractional signal modification of $6\mu$m wavelength hadron density perturbations after traversing a drift of specified length after the kicker and a bend within a dipole of specified angle. We find that moderate drift lengths will produce signal modification amplitudes of a few percent at the start of the dipole.}
\end{figure}

In the current design of the cooler, it is useful to have as much space as possible dedicated to the modulator, kicker, and amplifiers, making it difficult to also fit in an optimized transfer line between the kicker and detector to achieve optimal signal modification. We therefore consider the case where we simply have a drift after the kicker plus a main arc dipole (field strength of 3.782T, bending radius of 243m). We assume a hard-edge dipole model and that the beam path is normal to the pole face, so that edge effects may be ignored. We also use the fit wake of Fig.~\ref{fig:wake_fit_sat}. For convenience, we define the amplitude of signal modification so that positive values correspond to a reduction in intensity when the electrons and hadrons are aligned. We take an observation point at the start of the dipole and perform a scan of the amplitude of signal modification as a function of the $M_{56}$ value of a drift transfer matrix between the kicker center and the observation point for several wavelengths of hadron density perturbations, as shown in Fig.~\ref{fig:scan_wavelength_discrete}. We find an optimal wavelength of $\sim 6\mu$m, but the 1.2mm $M_{56}$ value corresponds to over 100m of drift, which is too long for the available space. We therefore focus on more reasonable parameters and scan the signal modification amplitude as a function of moderate kicker-to-dipole drift lengths and observation wavelengths, as shown in Fig.~\ref{fig:scan_wavelength}. Note that these drifts are defined between the end of the kicker and the start of the dipole, and so an extra $M_{56}$ contribution from half the kicker length is also included in the transfer matrix. We find an optimal observation wavelength of $6\mu$m over a range of drift lengths. We then fix a $6\mu$m observation wavelength and scan the signal modification amplitude as a function of drift length and bend angle within the dipole, as shown in Fig.~\ref{fig:scan_dipole}. We find that it is ideal to observe the radiation near the start of the dipole and that, within the region of interest, increased drift lengths result in increased signal modification, with amplitudes of a few percent.

\begin{figure}[!htbp]
\begin{center}
\includegraphics[width=1.0\columnwidth]{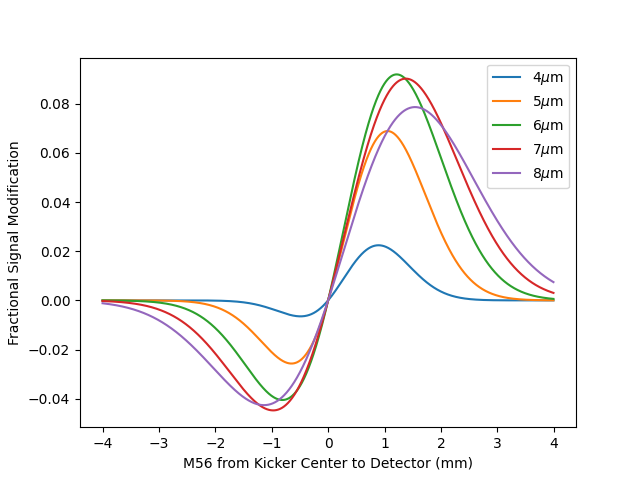}
\end{center}
\caption{\label{fig:scan_drift_negative} Fractional signal modification of various wavelengths as a function of $M_{56}$ between the kicker center and observation point, including negative values. We see signal modification with both signs of $M_{56}$.}
\end{figure}

\begin{figure}[!htbp]
\begin{center}
\includegraphics[width=1.0\columnwidth]{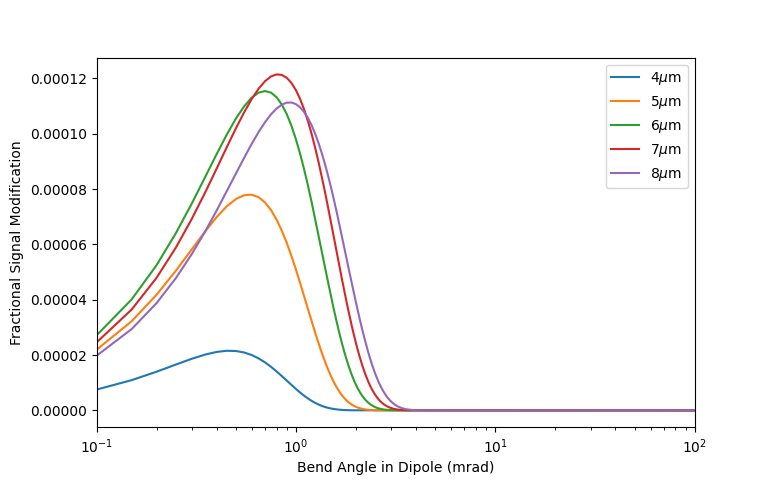}
\end{center}
\caption{\label{fig:scan_dipole_pure} Fractional signal modification of various wavelengths as a function of bend angle in the dipole, assuming only a real dipole between the center of the kicker and the observation point. The corresponding range of $M_{56}$ values is between -4cm and +9$\mu$m. The inclusion of non-zero $M_{51}$ and $M_{52}$ terms from the dipole destroys the signal modification. Note the horizontal log scale.}
\end{figure}

Note that the inclusion of the $M_{51}$ and $M_{52}$ terms is vital to the correct understanding of signal modification. Fig.~\ref{fig:scan_drift_negative} shows the result of scanning the amplitude of signal modification as a function of the $M_{56}$ term between the kicker center and observation point. It would appear that we could use a dipole to generate negative $M_{56}$ and see about half the signal modification we could obtain with a drift. However, the dipole will also generate non-trivial $M_{51}$ and $M_{52}$ terms. Performing a scan with the more realistic dipole generates the plot seen in Fig.~\ref{fig:scan_dipole_pure}. Virtually no signal modification is observed, and even then only for the small positive $M_{56}$ values generated near the very start of the dipole.

 \subsection{Fringe Fields}\label{subsec:fringe_fields}
 Since we are focused on the radiation at the start of the dipole, careful consideration of the fringe fields is important. We take a model of a dipole consisting of two magnetized poles of uniform magnetization $B_0/\mu_0$ separated by a distance $2w$ and of longitudinal length $L$ which extend infinitely far in the $x$ direction and are each semi-infinite in $y$. See Fig.~\ref{fig:dipole_geometry}. It can be analytically shown that, with the origin defined as the dipole center, the fields visible to the beam are given by:

\begin{figure}[!htbp]
\begin{center}
\includegraphics[width=1.0\columnwidth]{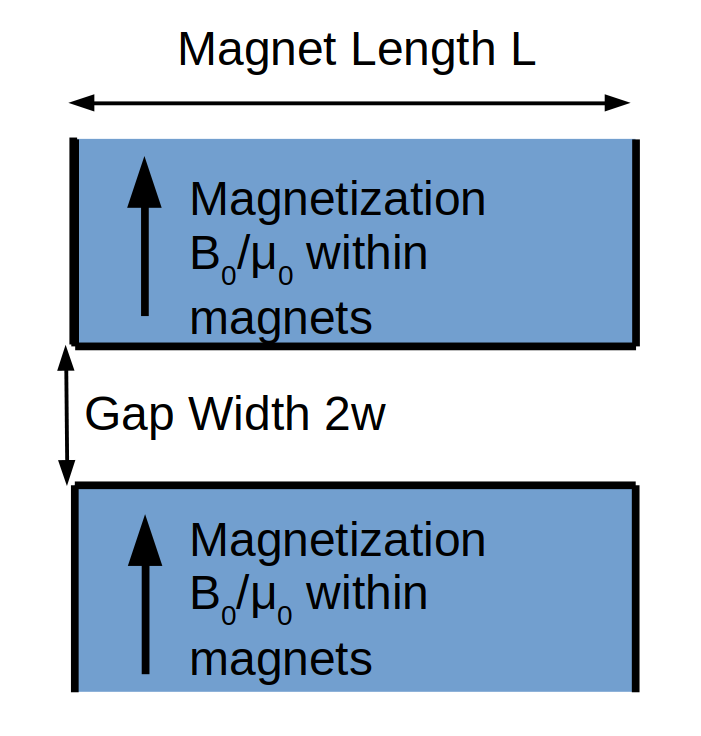}
\end{center}
\caption{\label{fig:dipole_geometry} Model of dipoles used to extract analytic fringe field expressions. The origin is at the center of the gap between the pole faces, with the $y$ axis pointing up and the the $z$ axis to the right. The poles are assumed infinite in $x$ and semi-infinite in $y$.}
\end{figure}

\begin{align}
B_y(y,z) = \frac{B_0}{2\pi}\Bigg[&\tan^{-1}\bigg(\frac{L/2+z}{w+y}\bigg)\\\nonumber
+ &\tan^{-1}\bigg(\frac{L/2-z}{w+y}\bigg)\\\nonumber
+&\tan^{-1}\bigg(\frac{L/2+z}{w-y}\bigg)\\\nonumber
+&\tan^{-1}\bigg(\frac{L/2-z}{w-y}\bigg)\Bigg]    
\end{align}

\begin{align}
    B_z(y,z) = \frac{B_0}{4\pi}\Bigg[&\ln\bigg(\frac{(w+y)^2 + (L/2+z)^2}{(w-y)^2 + (L/2+z)^2}\bigg)\\\nonumber
    +&\ln\bigg(\frac{(w-y)^2 + (L/2-z)^2}{(w+y)^2 + (L/2-z)^2}\bigg)\Bigg]
\end{align}

We choose the dimensions of the RHIC dipoles, with $w = 4$cm and $L=9.441$m. As before, $B_0 = 3.782$T \cite{cite:eic_cdr, cite:other_dipole_ref}. Although this is not an exact fit to any particular dipole, it provides a proof-of-principle study. In order to counteract the fringe fields, so that we arrive at a reasonable transfer matrix inside the main arc dipole where the field is strong enough to produce detectable radiation, we also include a 1m long screen dipole with $w=4$cm and a weak field opposite to that of the main arc dipole ($B_0 = -0.40$T).

\begin{figure}[!htbp]
\begin{center}
\includegraphics[width=1.0\columnwidth]{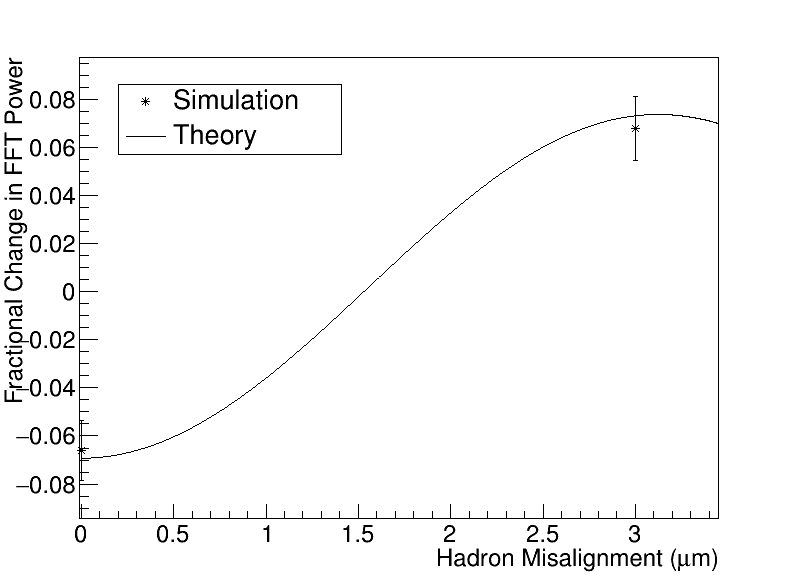}
\end{center}
\caption{\label{fig:sigsup_w_real_dipole} Fractional signal modification at a wavelength of $6.25\mu$m in theory and simulation for a 10 m drift after the end of the kicker plus a 0.14mrad bend in a 3.782T dipole. The simulation includes realistic fringe fields, partially corrected by a weak screen dipole to achieve the same total bend angle as expected in theory. Both theory and simulation make use of the idealized black wake in Fig.~\ref{fig:wakes_both}.}
\end{figure}

We check that this produces reasonable results by altering the simulation used in subsection \ref{subsec:sim_lin} to bring the hadrons from the kicker center to a point 5m after the end of the kicker through the use of a 27.5m drift transfer matrix, then track them with the realistic fields through a drift of 4m, the 1m screen dipole, and 91.5mm into the main arc dipole. The total integrated field seen by the proton making this journey is 0.13T-m, corresponding to a bend of 0.14 mrad. Due to the longer computation time, only 8 random seeds are used. We compare this to the theoretical prediction of a 32.5m drift after the kicker center (corresponding to 10m past the end of the kicker) plus a 0.14mrad bend in the main arc dipole. The results are shown in Fig.~\ref{fig:sigsup_w_real_dipole}. Good agreement is observed.

\subsection{Radiation}\label{subsec:radiation}

So far, we have focused entirely on the changes to the hadron density spectrum. However, this cannot be observed directly. Instead, we will monitor the radiation produced by the hadron beam as it moves through a dipole. In the far-field approximation, the electric field from an accelerated charge is given by \cite{cite:jackson}

\begin{align}\label{eqtn:jackson}
    \vec{E}(\vec{x},t_0) = \frac{q}{4\pi\epsilon_0c}\Bigg[\frac{1}{|\vec{r}_{0} - \vec{r}|}\frac{\hat{n}\times\big[(\hat{n}-\vec{\beta})\times\dot{\vec{\beta}}\big]}{\big(1-\vec{\beta}\cdot\hat{n}\big)^3}\Bigg]_t
\end{align}
\noindent
where the observation point is $\vec{r}_0$, the observation time is $t_0$, and the hadron position, $\vec{r}$, relativistic beta, $\vec{\beta}$, and its derivative, $\dot{\vec{\beta}}$, are evaluated at the retarded time, $t$, with $t_0 = t + |\vec{r}_{0} - \vec{r}|/c$.

Since we had seen in subsection \ref{subsec:dipole_obs} that it is useful to observe the signal modification immediately at the start of the dipole, we consider the edge radiation \cite{cite:edge_radiation}. We write the frequency-space electric field

\begin{align}\label{eqtn:fft_E}
    \vec{E}(\vec{x},\omega) &= \int_{-\infty}^{\infty} \vec{E}(\vec{x},t_0) e^{i\omega t_0} dt_0\\\nonumber
    &= \frac{q}{4\pi\epsilon_0c}\int_{-\infty}^{\infty}\Bigg[\frac{1}{|\vec{r}_{0} - \vec{r}|}\frac{\hat{n}\times\big[(\hat{n}-\vec{\beta})\times\dot{\vec{\beta}}\big]}{\big(1-\vec{\beta}\cdot\hat{n}\big)^2}\Bigg]_t e^{i\omega t_0} dt
\end{align}
\noindent
where we have simplified using the fact that $d t_0/dt = 1 - \vec{\beta}\cdot\hat{n}$.

The intensity of the radiation is given by \cite{cite:edge_radiation}

\begin{align}\label{eqtn:intensity}
    S = \alpha \frac{\Delta \omega}{\omega} \frac{I_h}{q} \bigg(\frac{2\epsilon_0c}{e}\bigg)^2\hbar\omega\Big|\vec{E}(\vec{x},\omega)\Big|^2
\end{align}
\noindent
where $\alpha$ is the fine-structure constant, $\Delta\omega$ is the bandwidth, and $I_h$ is the average hadron beam current (1A for the EIC).

\begin{figure}[!htbp]
\begin{center}
\includegraphics[width=1.0\columnwidth]{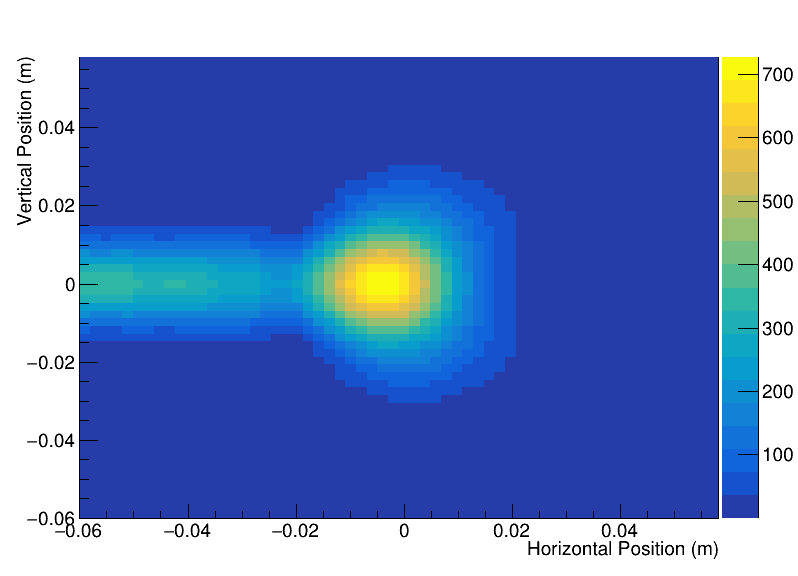}
\end{center}
\caption{\label{fig:radiation_detection_scan} Intensity of radiation from all protons in the beam seen 10m downstream of the dipole entrance within a 10\% bandwidth. Units are $\mu$W/m$^2$.}
\end{figure}

We track a test charge through the dipole fields described in subsection \ref{subsec:fringe_fields}, and use its path to integrate Eq.~\ref{eqtn:fft_E} at various locations on a transverse grid 10m downstream of the entrance pole face of the main arc dipole, where we might place a camera to detect the emitted radiation. We assume a 10\% bandwidth for this detector. The result of this is shown in Fig.~\ref{fig:radiation_detection_scan}. We see that we may achieve intensities of $\sim700\mu W/m^2$.

\begin{figure}[!htbp]
\begin{center}
\includegraphics[width=1.0\columnwidth]{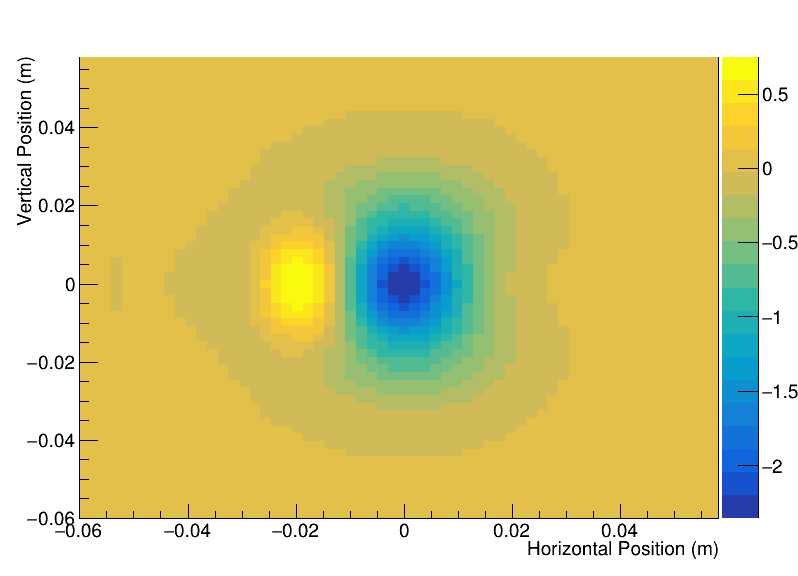}
\end{center}
\caption{\label{fig:radiation_intensity_change} Change in radiation intensity at a detector 10m downstream of the dipole entrance expected from signal modification with no hadron/electron misalignment. Units are $\mu$W/m$^2$. We see $\sim0.3\%$ reduction in intensity.}
\end{figure}

We repeat the above procedure, but model the effect of signal modification by multiplying the field amplitude produced by each step of the tracker by $\sqrt{1 + A(s)\sigma_{z,e}/\sigma_{z,h}}$, where $A(s)$ is the amplitude of signal suppression from Eq.~\ref{eqtn:pwr_relative}. We assume $6\mu$m light and use the fit wake of Fig.~\ref{fig:wake_fit_sat} (which includes saturation effects) and a transfer matrix through the relevant drift and concatenated dipole fields to reach an arbitrary location after the kicker. We assume 10m between the end the the kicker and the start of the main arc dipole, with a 1m screen dipole immediately in front of the main arc dipole. $\sigma_{z,e}/\sigma_{z,h}$ is the ratio of the electron to hadron bunch lengths, since only the central hadrons will see the effect of the cooling. We subtract the intensity shown in Fig.~\ref{fig:radiation_detection_scan} from that of this modified simulation in order to obtain the intensity of our signal. This is shown in Fig.~\ref{fig:radiation_intensity_change}. The fractional change is $\sim0.3\%$. Integration over 1mm$^2$ for 1 second on-axis will provide $2.0\times10^{10}$ photons from the full bunch, while signal modification in the core protons will reduce this number by $7.1\times10^7$. The error on the former should follow Poisson statistics, with a numerical value of $\sqrt{2.0\times10^{10}} \approx 1.4\times10^5$ so that we obtain a signal-to-noise ratio of $\sim 500$. However, inefficiencies in the detector will reduce this, so that we may need to integrate for a longer time or over a wider area.

We must also contend with background thermal radiation, whose intensity is given by the well-known blackbody radiation formula:

\begin{align}\label{eqtn:thermal}
    S = \frac{\hbar}{4\pi^2c^2}\frac{\omega^3}{e^{\hbar\omega/k_BT}-1} \Delta \omega
\end{align}

Within this same 10\% bandwidth at a temperature of 300K, we have a thermal intensity of $10W/m^2$, which would swamp our signal. It is therefore necessary to operate at liquid nitrogen temperature (77.29K), where the thermal background is only $1nW/m^2$.

\section{Conclusion}\label{sec:conclude}

We have derived an expression for signal modification for the case of a CEC cooler. We have also performed simulations of this process, finding good agreement with theoretical expectations. Finally, we have shown that such a signal may be experimentally observed at the few parts-per-thousand level with a cryo-cooled infrared camera set up to observe existing dipole radiation. As the design of the cooler matures, it will be necessary to incorporate the needs of this radiation diagnostic, at least ensuring the existence of a sufficient drift length between the end of the kicker and the first downstream dipole. We will also investigate the possibility of separating the radiation from the core protons from that of the rest of the bunch, which would increase the fractional change in radiation power by a factor of nearly 10, corresponding to the the ratio of hadron to electron bunch lengths. This may require operation at a shorter wavelength. Although we have provided a compelling proof-of-principle argument, we will also need to more carefully consider the details of the fringe fields of the dipole and appropriate compensation.

\section{Acknowledgements}\label{sec:ack}

This work was supported by Brookhaven Science Associates, LLC under Contract No. DE-SC0012704 with the U.S. Department of Energy, and by the Department of Energy, contract DE-AC03-76SF00515. We would like to thank the members of the EIC strong hadron cooling group for useful discussions.

\appendix 

\section{Alternative Derivations of Signal Modification}\label{app:alternative}

We present two alternative derivations of the theory of signal modification in the purely longitudinal case. In subsection \ref{subsec:thry_mike}, we adopt a hybrid approach where we integrate the Vlasov equation in the time domain and make use of single-particle statistics. In subsection \ref{subsec:thry_stupakov}, we back-propagate the phase-space density from the detector to the modulator, in line with the conventions of \cite{cite:stupakov_initial}. In the limit where the hadron delay depends only on its energy offset, ie $M_{51} = M_{52} = M_{53} = M_{54} = 0$, these results agree with Eq.~\ref{eqtn:pwr_relative}.

\subsection{Integration of Time-Domain Vlasov Equation}\label{subsec:thry_mike}

We consider the hadron beam entering the modulator as a constant background current, $I_0$, with a random Schottky signal, $I_s(s,t)$, overlaid on top, where $s$ is the longitudinal position within the cooler and $t$ is the time of measurement. After passage through the kicker, the beam will pick up a coherent response current, $I_c(s,t)$, due to the action of the wake, so that the total beam current at position $s$ and time $t$ is given by

\begin{align}\label{eqtn:mike_current}
    I(s,t) = I_0 + I_s(s,t) + I_c(s,t)
\end{align}

Since we assume only random noise in the hadron beam when it enters the modulator, we take $I_c(s,t) = 0$ for $s < s_K$, the kicker location. We also consider the respective phase-space densities, $\psi(\eta,s,t)$, of these three currents, with 

\begin{align}\label{eqtn:mike_phis}
    I(s,t) = q\int_{-\infty}^{\infty} \dot{s}(\eta,s) \psi(\eta, s, t) d\eta
\end{align}
\noindent
where $\dot{s}(\eta,s)$ is the local speed of the hadrons as a function of fractional energy deviation, $\eta$, and longitudinal position in the cooler.

The fluctuating part of the Schottky current can be written as

\begin{align}\label{eqtn:mike_schottky}
    I_s(s,t) = \sum_{k=-\infty}^{\infty}q \delta(t - t_k - \tau_0(s) - \eta_k\tau_1(s)) - I_0
\end{align}
\noindent
where $t_k$ is the time that particle $k$ arrives at the modulator, at position $s=0$, $\tau_0(s)$ is the time for an on-energy particle to travel from the modulator to point $s$, $\tau_1(s)$ is the energy-dependence of this transit time, and $\eta_k$ is the fractional energy deviation of particle $k$.

We define a frequency-space version of the Schottky current, $\tilde{I}_{s}(s, \omega_m)$, with the forward and inverse Fourier transforms

\begin{align}\label{eqtn:mike_fourier}
    \tilde{I}_{s}(s,\omega_m) &= \frac{1}{T}\int_{\tau_0(s)-T/2}^{\tau_0(s)+T/2} e^{i\omega_mt}I_s(s,t)dt\\\nonumber
    I_s(s,t) &= \sum_{m=-\infty}^{\infty}e^{-i\omega_m t}\tilde{I}_{s}(s,\omega_m)
\end{align}
\noindent
where $T$ is chosen to be large compared to typical values of $\eta_k\tau_1(s)$ and the wake wavelength divided by $\beta c$, so that we may ignore the precise behavior at the endpoints, and where $\omega_m = 2\pi m/T$. Particular particles near the edges of this time interval may enter or leave it during the passage through the cooler, but the relatively large size of $T$ ensures that they are a negligible fraction of the total, and can be safely ignored. If we assume $N$ particles arriving in the cooling system within time $T$ at times $t_a$, we obtain for $m \neq 0$

\begin{align}\label{eqtn:mike_I_sm}
    \tilde{I}_{s}(s, \omega_m) = \frac{q}{T}\sum_{a=1}^{N} e^{i\omega_m[t_a + \tau_0(s) + \eta_a\tau_1(s)]}
\end{align}

In the kicker, each particle receives a kick, defined by a time-dependent voltage

\begin{align}\label{eqtn:mike_voltage}
    V(t) &= \int_{-\infty}^{\infty}w_t(t - \tau - \hat{\tau}_0)I_s(0, \tau)d\tau
\end{align}
\noindent
where $\hat{\tau}_0$ is the delay of the electron beam in travelling between modulator and kicker and $w_t(t)$ is related to the wake function of Eq.~\ref{eqtn:wake_def} by
\begin{align}\label{eqtn:def_wake_time}
    w_t(t) = w(-\beta c t)
\end{align}
\noindent
and has an impedance, $Z_t(\omega)$, related to the impedance of Eq.~\ref{eqtn:imped_def} by
\begin{align}\label{eqtn:def_imped_time}
    Z_t(\omega) = \int_{-\infty}^{\infty}w_t(t)e^{i\omega t}dt = -\frac{Z(\omega/\beta c)}{\beta}
\end{align}
\noindent

After passing through the kicker, the hadron phase-space density also includes a perturbation described by the coherent response function, obeying the linearized Vlasov equation:
\begin{align}\label{eqtn:mike_vlasov}
    0 = &\frac{\partial\psi_c(\eta,s,t)}{\partial t} + \frac{\partial}{\partial s}(\dot{s}(\eta,s)\psi_c(\eta,s,t))\\\nonumber
    &+ \frac{qV(t)\delta(s-s_K)\dot{s}(\eta,s)}{E_0}\frac{\partial}{\partial \eta}(\psi_0(\eta))
\end{align}
\noindent
where we have explicitly written the $\dot{\eta}$ as a delta function kick from the kicker voltage. We have also pulled the factor of $\dot{s}$ out of the derivative with respect to energy since we operate in the ultra-relativistic regime and so assume that $\dot{s}$ is independent of particle energy within the kicker straight section.

Since there is no coherent response for $s< s_K$, integrating the above equation from $s_K^-$ to $s_K^+$ yields

\begin{align}\label{eqtn:mike_psic_real}
    \psi_c(\eta,s_K^+,t) = -\frac{qV(t)}{E_0}\frac{d\psi_0(\eta)}{d\eta}
\end{align}
\noindent
which may be transformed to frequency space as
\begin{align}\label{eqtn:mike_psic_fourier}
    \tilde{\psi}_{c}(\eta,s_K^+,\omega_m) = -\frac{q\tilde{V}(\omega_m)}{E_0}\frac{d\psi_0(\eta)}{d\eta}
\end{align}
\noindent
where we have defined a frequency-space voltage, $\tilde{V}(\omega_m)$, and phase-space density, $\tilde{\psi}_c(\eta,s,\omega_m)$, using the same procedure as in Eq.~\ref{eqtn:mike_fourier}.

Integration of Eq.~\ref{eqtn:mike_vlasov} to arbitrary $s > s_K$ yields

\begin{align}\label{eqtn:mike_int_psi}
    \tilde{\psi}_{c}(\eta,s,\omega_m) &= \tilde{\psi}_{c}(\eta,s_K^+,\omega_m)\\\nonumber
    &\times e^{i\omega_m[\tau_0(s) + \eta \tau_1(s) - \tau_0(s_K) - \eta \tau_1(s_K)]}
\end{align}

Integrating over particle energy to get the coherent current response, we find
\begin{align}\label{eqtn:mike_Ic}
    &\tilde{I}_{c}(s,\omega_m) = -\frac{q^2\tilde{V}(\omega_m)\dot{s}_0}{E_0}\\\nonumber
    &\times\int_{-\infty}^{\infty} \frac{d\psi_0(\eta)}{d\eta}e^{i\omega_m[\tau_0(s) + \eta \tau_1(s) - \tau_0(s_K) - \eta \tau_1(s_K)]} d\eta\\\nonumber
    &=\frac{q^2\tilde{V}(\omega_m)\dot{s}_0}{E_0}i\omega_m[\tau_1(s) - \tau_1(s_K)]\\\nonumber
    &\times\int_{-\infty}^{\infty} \psi_0(\eta)e^{i\omega_m[\tau_0(s) + \eta \tau_1(s) - \tau_0(s_K) - \eta \tau_1(s_K)]} d\eta\\\nonumber
    &=\tilde{I}_{s}(s=0,\omega_m)Z_t(\omega_m)e^{i\omega_m\hat{\tau}_0}\frac{q^2\dot{s}_0}{E_0}i\omega_m[\tau_1(s) - \tau_1(s_K)]\\\nonumber
    &\times\int_{-\infty}^{\infty} \psi_0(\eta)e^{i\omega_m[\tau_0(s) + \eta \tau_1(s) - \tau_0(s_K) - \eta \tau_1(s_K)]} d\eta
\end{align}
\noindent
where we have assumed that the change in $\dot{s}(\eta,s)$ due to the beam energy spread is small relative to the speed of an on-energy particle, $\dot{s}_0$, and made use of Eq.~\ref{eqtn:mike_I_sm}, \ref{eqtn:mike_voltage}, and \ref{eqtn:def_imped_time} to write the voltage in the kicker in terms of the frequency-domain Schottky current in the modulator and the impedance $Z_t(\omega_m)$. We use the above to define an effective gain function, $\tilde{I}_{c}(s,\omega_m) = G(s,\omega_m)\tilde{I}_{s}(s=0,\omega_m)$. For $m\neq0$, so that we may ignore the base current, $I_0$, the total current is

\begin{align}\label{eqtn:mike_It}
    \tilde{I}_{t}(s,\omega_m) = G(s,\omega_m)\tilde{I}_{s}(s=0,\omega_m) + \tilde{I}_{s}(s,\omega_m)
\end{align}
\noindent

We define the signal modification by computing 
\begin{align}\label{eqtn:mike_sigsup}
    \langle|\tilde{I}_{t}(s,\omega_m)|^2\rangle &= \langle|\tilde{I}_{s}(s,\omega_m)|^2\rangle (1 + |G(s,\omega_m)|^2)\\\nonumber
    &+ 2 Re \langle G(s,\omega_m)\tilde{I}_{s}(s=0,\omega_m)\tilde{I}^*_{s}(s,\omega_m)\rangle
\end{align}
\noindent
where the angle brackets denote the ensemble average and the star refers to complex conjugation.

Using Eq.~\ref{eqtn:mike_I_sm}, we can evaluate
\begin{align}\label{eqtn:mike_I_corr}
    \langle \tilde{I}_{s}&(s=0,\omega_m)\tilde{I}^*_{s}(s,\omega_m)\rangle\\\nonumber
    &=\frac{q^2}{T^2}\sum_{a=1}^N\sum_{b=1}^{N} e^{i\omega_m[t_a - t_b - \tau_0(s) - \eta_b\tau_1(s)]}\\\nonumber
    & = \frac{q^2}{T^2}N \Big\langle e^{-i\omega_m[\tau_0(s) + \eta\tau_1(s)]}\Big\rangle_{\eta}
\end{align}
\noindent
where we only kept the diagonal terms due to the uncorrelated nature of the initial particle distribution.

Similarly, we find
\begin{align}\label{eqtn:miked_self_corr}
    \langle |I_{s}(s,\omega_m)|^2\rangle = \frac{q^2}{T^2}N
\end{align}

We can now write the fractional signal modification as
\begin{align}\label{eqtn:mike_rel_sigsup}
    \frac{\langle|I_{t}(s,\omega_m)|^2\rangle}{\langle|I_{s}(s,\omega_m)|^2\rangle} &= 1 + |G(s,\omega_m)|^2\\\nonumber
    &+ 2\Re(Y(s,\omega_m))
\end{align}
\noindent
with

\begin{align}\label{eqtn:mike_g}
    G(s,\omega_m) &= iZ_t(\omega_m)\omega_m\frac{qI_0}{E_0}[\tau_1(s) - \tau_1(s_K)]\\\nonumber
    &\times e^{i\omega_m[\hat{\tau}_0 + \tau_0(s) - \tau_0(s_K)]}\Big\langle e^{i\omega_m\eta[\tau_1(s) - \tau_1(s_K)]} \Big\rangle_{\eta}
\end{align}
\noindent
and
\begin{align}\label{eqtn:mike_Y}
Y(s,\omega_m) = G(s,\omega_m) e^{-i\omega_m\tau_0(s)} \Big\langle e^{-i\omega_m\eta\tau_1(s)}\Big\rangle_{\eta}
\end{align}
\noindent
where the mean current $I_0 = q\dot{s}_0\int_{-\infty}^{\infty}\psi_0(\eta)d\eta$.

Since the kick from the wake is expected to be small, we ignore the second-order $|G(s,\omega_m)|^2$ term. We also assume a Gaussian energy distribution in the beam, with RMS energy error $\sigma_{\eta}$. This allows us to perform the energy averages explicitly, and we obtain the relative signal modification

\begin{align}\label{eqtn:mike_final}
\frac{\langle|I_{t}(s,\omega_m)|^2\rangle}{\langle|I_{s}(s,\omega_m)|^2\rangle} - 1 &= 2Re\Big(iZ_t(\omega_m)e^{i\omega_m[\hat{\tau}_0-\tau_0(s_K)]}\Big)\\\nonumber
&\times\omega_m\frac{qI_0}{E_0}[\tau_1(s) - \tau_1(s_K)]\\\nonumber
&\times e^{-\frac{\sigma_{\eta}^2\omega_m^2}{2}\big[\tau_1^2(s) + (\tau_1(s) - \tau_1(s_K))^2\big]}
\end{align}

We note that $\tau_1(s) = -1/\beta c$ times the $M_{56}$ element from the modulator to point $s$, $\hat{\tau}_0 - \tau_0(s_K) = \Delta z/\beta c$, and $I_0 = qn_0\beta c$, where $n_0$ is the mean longitudinal hadron density. We identify the wavenumber $k = \omega_m/\beta c$ and use Eq.~\ref{eqtn:def_imped_time} to translate $Z_t(\omega) \rightarrow -Z(k)/\beta$. This translates the fractional signal modification above to

\begin{align}\label{eqtn:mike_final_translated}
    &\frac{\langle|I_{t}(s,\beta c k)|^2\rangle}{\langle|I_{s}(s,\beta c k)|^2\rangle} - 1 = -2n_0\frac{q^2c}{E_0}k M_{56}^{KD}\\\nonumber
    &\times[\Im(Z(k))\cos(k\Delta z) + \Re(Z(k))\sin(k\Delta z)]\\\nonumber
&\times  e^{-\frac{\sigma_{\eta}^2k^2}{2}\Big[\big(M_{56}^{MD}\big)^2 + \big(M_{56}^{KD}\big)^2\Big]}
\end{align}

This agrees with Eq.~\ref{eqtn:pwr_relative} in the purely longitudinal limit.

\subsection{Theory from Back-Propagation of Phase Space Density}\label{subsec:thry_stupakov}

We now approach the problem of signal modification using the formalism described in \cite{cite:stupakov_initial}, except using the wake and impedance formalism from Eq.~\ref{eqtn:wake_def} and \ref{eqtn:imped_def}. This involves back-propagating the beam through the cooler to describe the phase space distribution of the beam at the detector in terms of the known distribution at the modulator. This gives us a longitudinal phase-space density at the detector equal to

\begin{align}\label{eqtn:stupakov_density}
f(z,\eta) = n_0F_0(&\eta - \Delta\eta(z - M_{56}^{KD}\eta))\\\nonumber
+ \delta f^{(m)}(&z - M_{56}^{MD}\eta - \Delta z + M_{56}^{MK}\Delta\eta(z - M_{56}^{KD}\eta),\\\nonumber
&\eta - \Delta \eta (z - M_{56}^{KD}\eta)) 
\end{align}
\noindent
where $F_0(\eta)$ is the energy-dependence of the base longitudinal phase-space density, $n_0$ is the mean longitudinal beam density, and $\delta f^{(m)}$ is some perturbation at the modulator. Note that $z$ and $\eta$ are evaluated at the detector, so that the above equation is just writing the original phase space density, $n_0F_0(\eta^{(m)}) + \delta f^{(m)}(z^{(m)}, \eta^{(m)})$, in terms of the detector $z$, $\eta$ coordinates.

We use Eq.~\ref{eqtn:wake_def} and \ref{eqtn:imped_def} to write

\begin{align}\label{eqtn:stupakov_kick}
    \Delta \eta(z) &= \frac{1}{2\pi}\int_{-\infty}^{\infty} e^{ikz} \Delta\tilde{\eta}(k) dk\\\nonumber
    &= -\frac{q^2c}{2\pi E_0}\int_{-\infty}^{\infty} e^{ikz} Z(k)\delta\tilde{n}^{(m)}_k dk
\end{align}
\noindent
where $Z(k)$ is the wake impedance and $\delta\tilde{n}_k^{(m)}$ is the Fourier transform of the hadron longitudinal density perturbation in the modulator, with a general definition

\begin{align}\label{eqtn:stupakov_n}
    \delta\tilde{n}_k &= \int_{-\infty}^{\infty} d\eta \delta \tilde{f}_k(\eta)\\\nonumber
    &= \int_{-\infty}^{\infty} d\eta \int_{-\infty}^{\infty} dz \delta f(z,\eta) e^{-ikz}
\end{align}

If we Taylor expand $F_0$ to first order about the energy kick from the kicker and ignore the $\Delta\eta$ dependence in the expression for $\delta f^{(m)}$, since it is already assumed to be a small correction to $F_0$, we find the perturbation to the phase space density at the detector:

\begin{align}\label{eqtn:stupakov_taylor}
    \delta f(z,\eta) &= -n_0F'_0(\eta)\Delta\eta(z - M_{56}^{KD}\eta)\\\nonumber
    &+ \delta f^{(m)}(z - M_{56}^{MD}\eta - \Delta z, \eta)
\end{align}

The power of signal modification is given by the correlator

\begin{align}\label{eqtn:stupakov_correlator}
    \langle\delta\tilde{n}_k \delta\tilde{n}_{k'}\rangle = \int_{-\infty}^{\infty} d\eta  d\eta' \langle \delta\tilde{f}_k(\eta) \delta\tilde{f}_{k'}(\eta')\rangle
\end{align}

We split the right hand side of Eq.~\ref{eqtn:stupakov_taylor} into two parts:
\begin{align}\label{eqtn:stupakov_split}
    &\delta f^{(1)}(z,\eta) \equiv -n_0F'_0(\eta)\Delta\eta(z - M_{56}^{KD}\eta)\\\nonumber
    &\delta f^{(2)}(z,\eta) \equiv \delta f^{(m)}(z - M_{56}^{MD}\eta - \Delta z, \eta)
\end{align}

Taking the Fourier transform of the first term, we find
\begin{align}\label{eqtn:stupkov_f1_transform}
    \delta \tilde{f}^{(1)}(k) &= -n_0F'_0(\eta)\int_{-\infty}^{\infty}e^{-ikz}\Delta\eta(z - M_{56}^{KD}\eta)dz\\\nonumber
    &=-n_0F'_0(\eta)e^{-ikM_{56}^{KD}\eta}\int_{-\infty}^{\infty}e^{-ikz'}\Delta\eta(z')dz'\\\nonumber
    &=n_0F'_0(\eta)\frac{q^2c}{E_0}Z(k)e^{-ikM_{56}^{KD}\eta}\delta\tilde{n}^{(m)}_k
\end{align}
\noindent
where we have made use of Eq.~\ref{eqtn:stupakov_kick} in the last step.

The Fourier transform of the second term in Eq.~\ref{eqtn:stupakov_split} yields
\begin{align}\label{eqtn:stupakov_f2_transform}
    \delta\tilde{f}^{(2)} &= \int_{-\infty}^{\infty} e^{-ikz} \delta f^{(m)}(z - M_{56}^{MD}\eta - \Delta z, \eta) dz\\\nonumber
    &= e^{-ikM_{56}^{MD}\eta}e^{-ik\Delta z}\int_{-\infty}^{\infty} e^{-ikz'} \delta f^{(m)}(z', \eta) dz'\\\nonumber
    &= e^{-ikM_{56}^{MD}\eta}e^{-ik\Delta z}\delta\tilde{f}^{(m)}_k(\eta)
\end{align}

Calculating the correlator in Eq.~\ref{eqtn:stupakov_correlator}, we see that it involves the sum of three terms: $\langle \delta\tilde{f}^{(1)}_k(\eta) \delta\tilde{f}^{(1)}_{k'}(\eta')\rangle$, $\langle \delta\tilde{f}^{(2)}_k(\eta) \delta\tilde{f}^{(2)}_{k'}(\eta')\rangle$, and $\langle \delta\tilde{f}^{(1)}_k(\eta) \delta\tilde{f}^{(2)}_{k'}(\eta') + \delta\tilde{f}^{(2)}_k(\eta) \delta\tilde{f}^{(1)}_{k'}(\eta')\rangle$. The first of these is quadratic in the wake impedance, and so we will ignore it since we assume that the kick from the wake is small. The second term is evaluated using Eq. 4 of \cite{cite:stupakov_initial}:
\begin{align}\label{eqtn:stupakov_self_corr}
    \langle \delta\tilde{f}^{(2)}_k(\eta) \delta\tilde{f}^{(2)}_{k'}(\eta')\rangle = 2\pi n_0F_0(\eta)\delta(k+k')\delta(\eta-\eta')
\end{align}
\noindent
and integration over energy deviations yields
\begin{align}\label{eqtn:stupakov_self_n}
    \langle\delta\tilde{n}_k^{(2)}\delta\tilde{n}_{k'}^{(2)}\rangle = 2\pi n_0\delta(k+k')
\end{align}

For the third term, $\langle \delta\tilde{f}^{(1)}_k(\eta) \delta\tilde{f}^{(2)}_{k'}(\eta') + \delta\tilde{f}^{(2)}_k(\eta) \delta\tilde{f}^{(1)}_{k'}(\eta')\rangle$, we note that the two parts become identical if we swap $k \leftrightarrow k'$. (We integrate over $\eta$ and $\eta'$, so the switch of these variables doesn't matter.) We then focus on the first of these

\begin{align}\label{eqtn:stupakov_term3}
    \int_{-\infty}^{\infty}d\eta d\eta'\langle \delta\tilde{f}^{(1)}_k(\eta) \delta&\tilde{f}^{(2)}_{k'}(\eta')\rangle\\\nonumber
    =n_0\frac{q^2c}{E_0}Z(k) \int_{-\infty}^{\infty} &F'_0(\eta)e^{-ikM_{56}^{KD}\eta}e^{-ik'M_{56}^{MD}\eta'}e^{-ik'\Delta z}\\\nonumber &\langle\delta\tilde{n}_k^{(m)}\delta\tilde{f}^{(m)}_{k'}(\eta')\rangle d\eta d\eta'
\end{align}

The correlator $\langle\delta\tilde{n}_k^{(m)}\delta\tilde{f}^{(m)}_{k'}(\eta')\rangle$ is the integral of Eq.~\ref{eqtn:stupakov_self_corr} over $\eta$, yielding
\begin{align}\label{eqtn:stupakov_term3b}
 &\int_{-\infty}^{\infty}d\eta d\eta'\langle \delta\tilde{f}^{(1)}_k(\eta) \delta\tilde{f}^{(2)}_{k'}(\eta')\rangle\\\nonumber
    =&2\pi n_0^2\frac{q^2c}{E_0}Z(k)\delta(k+k')\\\nonumber
    \times&\int_{-\infty}^{\infty}d\eta d\eta' F'_0(\eta)e^{-ikM_{56}^{KD}\eta}e^{-ik'M_{56}^{MD}\eta'}e^{-ik'\Delta z} F_0(\eta')\\\nonumber
    =&2\pi ikM_{56}^{KD} n_0^2\frac{q^2c}{E_0}Z(k)\delta(k+k')\\\nonumber
    \times&\int_{-\infty}^{\infty}d\eta d\eta' F_0(\eta)e^{-ikM_{56}^{KD}\eta}e^{ikM_{56}^{MD}\eta'}e^{ik\Delta z} F_0(\eta')
\end{align}

If we take the particular case of a Gaussian energy distribution with RMS $\sigma_{\eta}$, the integrals yield

\begin{align}\label{eqtn:stupakov_term3c}
 &\int_{-\infty}^{\infty}d\eta d\eta'\langle \delta\tilde{f}^{(1)}_k(\eta) \delta\tilde{f}^{(2)}_{k'}(\eta')\rangle\\\nonumber
    =&2\pi ikM_{56}^{KD} n_0^2\frac{q^2c}{E_0}Z(k)\delta(k+k')e^{ik\Delta z}\\\nonumber
    &\times e^{-\sigma_{\eta}^2k^2[(M_{56}^{KD})^2 + (M_{56}^{MD})^2]/2}
\end{align}

As noted above, the correlator corresponding to $\langle \delta\tilde{f}^{(2)}_k(\eta) \delta\tilde{f}^{(1)}_{k'}(\eta')\rangle$ is the same as what we have just computed, with the interchange $k \leftrightarrow k'$. Noting that the delta function sends $k' \leftrightarrow -k $ and that $Z(k)$ is the Fourier transform of a real-valued function, so that $Z(k')= Z(-k) = Z^*(k)$, we sum the two halves of this correlator to obtain
\begin{align}\label{eqtn:stupakov_term3_full}
    &2\pi ikM_{56}^{KD} n_0^2\frac{q^2c}{E_0}[Z(k)e^{ik\Delta z} - Z^*(k)e^{-ik\Delta z}]\\\nonumber
    &\times e^{-\sigma_{\eta}^2k^2[(M_{56}^{KD})^2 + (M_{56}^{MD})^2]/2}\delta(k+k')\\\nonumber
    =&-4\pi kM_{56}^{KD} n_0^2\frac{q^2c}{E_0}(Re[Z(k)]\sin(k\Delta z) + Im[Z(k)]\cos(k\Delta z))\\\nonumber
    &\times e^{-\sigma_{\eta}^2k^2[(M_{56}^{KD})^2 + (M_{56}^{MD})^2]/2}\delta(k+k')
\end{align}

Adding in the result from Eq.~\ref{eqtn:stupakov_self_n}
\begin{align}\label{eqtn:stupakov_final}
\langle\delta\tilde{n}_k \delta\tilde{n}_{k'}\rangle &= 2\pi n_0\delta(k+k')\Big[1 - 2 kM_{56}^{KD} n_0\frac{q^2c}{E_0}\\\nonumber
&\times(Re[Z(k)]\sin(k\Delta z) + Im[Z(k)]\cos(k\Delta z))\\\nonumber
&\times e^{-\sigma_{\eta}^2k^2[(M_{56}^{KD})^2 + (M_{56}^{MD})^2]/2}\Big]
\end{align}

The fractional signal modification is then
\begin{align}\label{eqtn:stupakov_final_frac}
\frac{\langle\delta\tilde{n}_k \delta\tilde{n}_{k'}\rangle}{\langle\delta\tilde{n}^{(m)}_k \delta\tilde{n}^{(m)}_{k'}\rangle} &- 1 = -2 kM_{56}^{KD} n_0\frac{q^2c}{E_0}\\\nonumber
&\times(Re[Z(k)]\sin(k\Delta z) + Im[Z(k)]\cos(k\Delta z))\\\nonumber
&\times e^{-\sigma_{\eta}^2k^2[(M_{56}^{KD})^2 + (M_{56}^{MD})^2]/2}
\end{align}

We see that, in the absence of transverse motion, this is identical to Eq.~\ref{eqtn:pwr_relative}.

\section{Quadratic Term}\label{app:second_order}

In the simplification of Eq.~\ref{eqtn:pwr}, we only kept the terms linear in the wake amplitude. If we instead keep terms to second order, we arrive at a corrective term equal to 
\begin{widetext}
\begin{align}\label{eqtn:second_order_sum}
\bigg(k M_{56}^{KD}\frac{q^2}{E_0}\bigg)^2 \sum_{i \neq a}&e^{-ik\big[z_m^{(i)} - z_m^{(a)} + M^{MD}_{5u}\big(\vec{x}_u^{(i)} - \vec{x}_u^{(a)}\big)\big]}\\\nonumber
\times \sum_{j,\ell} \bigg[&-\frac{1}{2}w(z_m^{(i)} + M_{5u}^{MK}\vec{x}_u^{(i)} + \Delta z - z_m^{(j)})w(z_m^{(i)} + M_{5u}^{MK}\vec{x}_u^{(i)} + \Delta z - z_m^{(\ell)})\\\nonumber
&-\frac{1}{2}w(z_m^{(a)} + M_{5u}^{MK}\vec{x}_u^{(a)} + \Delta z - z_m^{(j)})w(z_m^{(a)} + M_{5u}^{MK}\vec{x}_u^{(a)} + \Delta z - z_m^{(\ell)})\\\nonumber
&+w(z_m^{(i)} + M_{5u}^{MK}\vec{x}_u^{(i)} + \Delta z - z_m^{(j)})w(z_m^{(a)} + M_{5u}^{MK}\vec{x}_u^{(a)} + \Delta z - z_m^{(\ell)})\bigg]
\end{align}
\end{widetext}

If the integral of the wake is zero, taking the average of this quantity over all possible $z_m^{(j)}$ will yield zero unless $j=i$, $a$, or $\ell$. Similarly, the only values of $\ell$ which yield non-trivial results are $i$, $a$, or $j$. Since there are 4 ways to pick $j$ and $\ell$ equal to some combination of $i$ and $a$ and $\sim N>>4$ ways to pick them equal to one another and not equal to $i$ or $a$, we only focus on the latter case. Of the three wake cross-terms, only the last one survives, since the others have either $z_m^{(a)}$ or $z_m{(i)}$ appearing only in the complex phase, which would average to zero.

Doing the sums over the three free parameters, $i$, $a$, and $j$, we find this is equal to
\begin{align}\label{eqtn:second_order_int}
&N^3\bigg(k M_{56}^{KD}\frac{q^2}{E_0}\bigg)^2 \int_{-L/2}^{L/2} dz_m^{(i)}dz_m^{(a)}dz_m^{(j)}/L^3\\\nonumber
&\times\int_{-\infty}^{\infty}d^5\vec{x}^{(i)}d^5\vec{x}^{(a)}\rho\big(\vec{x}^{(i)}\big)\rho\big(\vec{x}^{(a)}\big) \\\nonumber
&\times e^{-ik\big[z_m^{(i)} - z_m^{(a)} + M^{MD}_{5u}\big(\vec{x}_u^{(i)} - \vec{x}_u^{(a)}\big)\big]}\\\nonumber
&\times w(z_m^{(i)} + M_{5u}^{MK}\vec{x}_u^{(i)} + \Delta z - z_m^{(j)})\\\nonumber
&\times w(z_m^{(a)} + M_{5u}^{MK}\vec{x}_u^{(a)} + \Delta z - z_m^{(j)})
\end{align}

Approximating the longitudinal integrals over $z_m^{(i)}$ and $z_m^{(a)}$ as extending from $-\infty$ to $\infty$, making the change of variables $z' \equiv z_m^{(i)} + M_{5u}^{MK}\vec{x}_u^{(i)} + \Delta z - z_m^{(j)}$ and $z'' \equiv z_m^{(a)} + M_{5u}^{MK}\vec{x}_u^{(a)} + \Delta z - z_m^{(j)}$, and using Eq.~\ref{eqtn:imped_def}, we obtain

\begin{align}\label{eqtn:second_order_int2}
&N\bigg(n_0 k M_{56}^{KD}\frac{q^2c}{E_0}\bigg)^2 \int_{-L/2}^{L/2} dz_m^{(j)}/L\\\nonumber
&\times\int_{-\infty}^{\infty}d^5\vec{x}^{(i)}d^5\vec{x}^{(a)}\rho\big(\vec{x}^{(i)}\big)\rho\big(\vec{x}^{(a)}\big) \\\nonumber
&\times e^{-ik\big[\big(M^{MD}_{5u} - M^{MK}_{56}\big)\big(\vec{x}_u^{(i)} - \vec{x}_u^{(a)}\big)\big]}Z(k)Z(-k)\\\nonumber
&=N\bigg(n_0 k M_{56}^{KD}\frac{q^2c}{E_0}\bigg)^2 |Z(k)|^2\\\nonumber
&\times\int_{-\infty}^{\infty}d^5\vec{x}^{(i)}d^5\vec{x}^{(a)}\rho\big(\vec{x}^{(i)}\big)\rho\big(\vec{x}^{(a)}\big) \\\nonumber
&\times e^{-ik\big[\big(M^{MD}_{5u} - M^{MK}_{56}\big)\big(\vec{x}_u^{(i)} - \vec{x}_u^{(a)}\big)\big]}\\\nonumber
&=N\bigg(n_0 k M_{56}^{KD}\frac{q^2c}{E_0}\bigg)^2 |Z(k)|^2e^{-k^2 \sum_{u\neq5} \sigma^2_{\hat{x}_u} \big(\hat{M}^{MD}_{5u} - \hat{M}^{Mk}_{5u}\big)^2}
\end{align}
\noindent
where we have used the fact that $Z(-k) = Z^*(k)$ for real wakes, the definition $n_0 = N/L$, and the transformed transfer matrices and beam sizes of Eq.~\ref{eqtn:m_tilde} - \ref{eqtn:sigmas}. This yields a corrected version of Eq.~\ref{eqtn:pwr_relative}:

\begin{align}\label{eqtn:pwr_relative_correct}
\frac{\Delta |\tilde{n}(k)|^2}{|\tilde{n}(k)|^2} &= -2n_0\frac{q^2c}{E_0}kM_{56}^{KD}\\\nonumber
&\times[\Re(Z(k))\sin(k\Delta z) + \Im(Z(k))\cos(k\Delta z)]\\\nonumber
&\times e^{-\frac{k^2}{2} \sum_{u\neq5} \sigma^2_{\hat{x}_u}\Big[\big(\hat{M}^{MD}_{5u}\big)^2 + \big(\hat{M}^{MD}_{5u} - \hat{M}^{MK}_{5u}\big)^2\Big]}\\\nonumber
&+\bigg(n_0 k M_{56}^{KD}\frac{q^2c}{E_0}\bigg)^2 |Z(k)|^2e^{-k^2 \sum_{u\neq5} \sigma^2_{\hat{x}_u} \big(\hat{M}^{MD}_{5u} - \hat{M}^{MK}_{5u}\big)^2}
\end{align}

Similar results are seen in the derivations of Appendix \ref{app:alternative} if we keep the $|G|^2$ term of Eq.~\ref{eqtn:mike_rel_sigsup} or the $\langle \delta\tilde{f}^{(1)}_k(\eta) \delta\tilde{f}^{(1)}_{k'}(\eta')\rangle$ term of Appendix \ref{subsec:thry_stupakov}.

\bibliography{sigsup_prab}

\end{document}